\documentclass[12pt]{article}
\usepackage{epsfig}
\usepackage{graphicx}
\usepackage{citesort}

\begin{document}

\title{Clustering Instabilities, Arching, and Anomalous \\
       Interaction Probabilities as Examples for \\
       Cooperative Phenomena in Dry Granular Media}
\author{S.~Luding\\
Institute for Computer Applications 1, \\
Pfaffenwaldring 27, D-70569 Stuttgart \\
GERMANY\\
e-mail: lui@ica1.uni-stuttgart.de
}
\date{in: T.A.S.K. Quarterly, Scientific Bulletin of Academic
          Computer Centre of the Technical University of Gdansk, 
          2(3), 417-443, July 1998.}

\maketitle

\section*{Abstract}
In a freely cooling granular material fluctuations in 
density and temperature cause position dependent energy loss.
Due to strong local dissipation, pressure and energy drop rapidly and 
material moves from `hot' to `cold' regions, leading 
to even stronger dissipation and thus causing the density 
instability. The assumption of `molecular chaos'
is valid only in the homogeneous cooling regime. As soon as the
density instability occurs, the impact parameter is not longer
uniformly distributed. The pair-correlation and the 
structure functions show that the molecular chaos assumption 
--- together with reasonable excluded volume modeling --- is
important for short distances and irrelevant on large 
length scales.

In this study, the probability distribution of the collision frequency
is examined for pipe flow and for freely cooling granular materials
as well. Uncorrelated events lead to a Poisson distribution 
for the collision frequencies.
In contrast, the fingerprint of the cooperative phenomena discussed here
is a power-law decay of the probability for many collisions per unit time.\\

\noindent {\bf Keywords:}
discrete element method, event driven simulations, clustering instability, 
arching, shock waves, power-law distribution, cooperative phenomena.

\newpage
\tableofcontents

\section{Introduction}

Many rather astonishing phenomena are known to occur
when granular materials like sand or powders move\ 
\cite{wolf97,behringer97,luding98,herrmann98}. Of interest are, 
e.g.~density waves emitted from outlets\ \cite{baxter89},
crack formation during vibration or during the flow through
a pipe \cite{duran94b,duran96,luding96,luding96b}, and
pattern formation due to dissipation 
\cite{goldhirsch93,melo94,luding96e,umbanhowar96,luding98}.
All these effects are connected to the ability of granular materials 
to form a hybrid state between a fluid and a solid: 
energy input can lead to a reduction of density so that the material 
becomes `fluid' and, on the other hand, in the
absence of energy input, granular materials `solidify' due to
dissipation. Thus, a packing of sand behaves like a solid when 
pushed, but offers no resistance to a pulling force.

In order to formalize and quantify the complicated rheology of granular 
media various attempts have been made. Continuum equations of motion and 
kinetic theories\ 
\cite{savage79,jenkins79,haff83,homsy92,hwang95,goldshtein95,herrmann98,luding98c}
are the first successful steps towards a quantitative description of
granular materials --- at least for limited parameter range.
This restriction exists because it is very difficult to incorporate into these 
theories all the details of static friction or other relevant microscopic mechanisms.
Also the generalization to high densities is an ardous task,
see for example Refs. \cite{brey96,dufty96} and references therein.
Most of the classical, but also the more advanced theories are based on the
assumption of molecular chaos --- the assumption that the velocities 
and the relative positions of all colliding pairs of particles in a gas
are uncorrelated.  In a dilute gas, the errors introduced by this assumption are
negligible.  In dense granular flows, however, correlations between colliding
particles may be important, leading to qualitative changes of behavior.

Section \ref{sec:methods} is dedicated to a 
brief introduction of the different discrete modeling approaches. 
In particular, we will present the `hard-particle' Event Driven (ED) 
\cite{allen87,lubachevsky91} and the `soft-particle' Molecular 
Dynamics (MD)\ \cite{cundall79,allen87,rapaport95} methods,
both for inelastic spherical particles with frictional forces. 
We extend the traditional ED method by introducing a cut-off time $t_c$
for dissipation \cite{luding96e,luding97c,luding98f}. Any particle that 
encounters a second collision before this time passed by is assumed
to be elastic --- the extended method is named TCED \cite{luding98f}.
Furthermore, the Direct Simulation Monte Carlo (DSMC) approach is
discussed and applied to free\-ly cooling granular media.
The validity of the molecular chaos assumption in granular flows
was examined by comparing event driven (ED) `hard sphere' simulations
to those performed with the Direct Simulation Monte Carlo (DSMC) method
\cite{luding98}. The ED method is capable of reproducing velocity correlations
--- even in the limit of rather large densities ---
whereas DSMC assumes molecular chaos by construction. In Section\ 
\ref{sec:EDvsDSMC} the structure factor and the pair-correlation function 
are examined and reasons for the breakdown of molecular chaos are
discussed. 
The clustering instability is described in section \ref{sec:cluster} 
with respect to the restitution coefficient and a cut-off time for dissipation.
The probability distribution function for the collision frequency is
measured in the homogeneous and the non-homogeneous clustering case
and differences are evidenced \cite{luding98}. 
The same is also found in pipe flow \cite{luding96b} where
shock waves and arching are the observed cooperative phenomena.
The probability distribution functions are measured and discussed
in Section\ \ref{sec:pipeflow} and interestingly have the same functional
behavior as in the case of the clustering instability. 
Finally, the results are summarized in Section\ \ref{sec:summary}.

\section{Models for particle-particle interactions}
\label{sec:methods}

The basic constitutents of granular materials are mesoscopic grains,
made of, for example, $10^{20}$ molecules. When these objects
interact (collide) the attractive potentials of the individual atoms
can often be neglected. Three models for the particle-particle
interactions are discussed in the following. They
account for the excluded volume of the particles via
a repulsive potential, either `hard' or `soft', or assume point
particles and introduce appropriate corrections.

It is important that the surface of the grains is rough on a microscopic 
scale so that solid friction occurs. In general, one has to distinguish
between sliding, sticking, and rolling friction, but we will
only discuss simplified models here. An entire discipline called tribology 
has evolved to study solid friction in depth\ \cite{johnson89,wolf97}. 
Friction and other sources of dissipation, like viscous damping or 
plastic deformations, have the crucial consequence that the system {\it does 
not conserve energy}. 
Since dissipation may occur due to various reasons, we discuss in the
following only simple dissipation laws, assuming that the detailed
knowledge of the interaction potential is of minor importance. In fact,
complicated laws often increase the number of parameters without giving
qualitatively different results \cite{luding94c}.

The difference between the two most frequently used discrete element
methods is the repulsive interaction potential. For the molecular dynamics (MD) 
method, soft particles with a power-law interaction potential are
assumed, whereas for the event driven (ED) method perfectly rigid particles
are used. The consequence is that the duration of the contact of two particles, 
$t_c$, is finite for MD, but vanishes for ED. In the DSMC method, one
assumes point particles without repulsive potential but with an effective 
scattering cross section. In addition one applies corrections from the kinetic
theory, in order to account for the effective free volume, the reduced mean 
free path, the increased collision frequency, and the enhanced collisional
momentum transport.

\subsection{The event driven, rigid particle method}

Here, we apply
the simplified collision model introduced by Walton and Braun 
\cite{walton86} and recently experimentally established by Foerster 
et al.\ \cite {foerster94} and Labous and Rosato \cite{labous97}. 
For given velocities before contact, three coefficients are needed
to evaluate the velocities after collision. At first, the
coefficient of normal restitution, $r$, defines the 
incomplete restitution of the normal component of the 
relative velocity. The second, the coefficient of friction, 
$\mu$, relates the tangential momentum change to the normal one, 
i.e.~Coulomb's law. The third, the coefficient
of maximum tangential restitution, $\beta_0$, delimits
the restitution of tangential velocity of the contact point 
to ensure energy conservation.  Note that this model implies that 
two grains at contact either slide, following Coulomb's law,
or stick together \cite {walton86,foerster94,luding95b,luding98b}. 
In the following, we apply the basic conservation laws
and determine the equations for the velocities after collision.

\begin{figure}
\begin{center}
\epsfig{figure=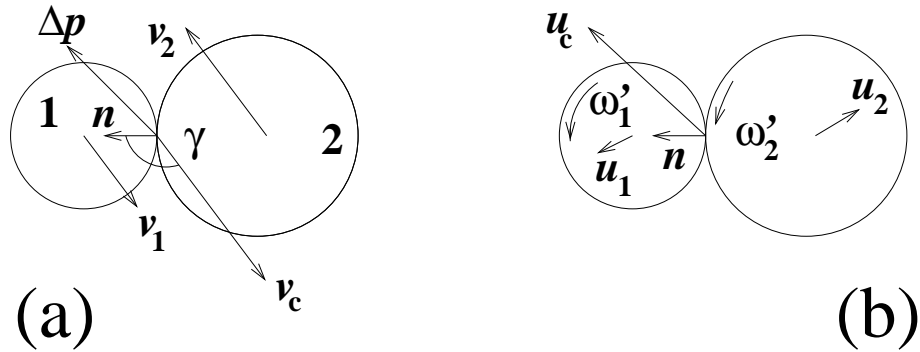,height=5.0cm}
\end{center}
\vspace{-1.8cm}~\\
\caption{Typical velocities of two particles immediately before (a) and
after (b) collision.}
\label{fig:coll_schem}
\end{figure}

Consider two particles with diameter $d_1$ and $d_2$ and masses $m_1$ and $%
m_2$. The normal unit vector for their contact is 
$\vec{n}={{{(\vec{r}_1-\vec{r}_2)\,}}/{{\,|\vec{r}_1-\vec{r}_2|}}}$, 
where $\vec{r}_i$ is the vector that gives the position of
the center of particle $i$ ($i$ = 1, 2). For the interaction of particle
$i=1$ with a fixed wall, we set $m_2=\infty $, and $\vec{n}$ is in 
this case the unit vector perpendicular to the wall surface pointing 
from the contact point with the wall to the center of the particle. 
The relative velocity of the contact points is 
\begin{equation}
\vec{v}_c=\vec{v}_1-\vec{v}_2-
          \left( {{\frac{d_1}2}\vec{\omega}_1+
                 {\frac{d_2}2}\vec{\omega}_2} \right) 
                                              \times \vec{n}~,
\end{equation}
with $\vec{v}_i$ and $\vec{\omega}_i$ being the
linear and angular velocities of particle $i$ just before collision. From
the momentum conservation laws for linear and angular momentum follows
\begin{equation}
\label{eq:uw}
\vec{v}_1^{\prime }=\vec{v}_1+\Delta \vec{P}/m_1~  \label{equationa}
\end{equation}
\begin{equation}
\vec{\omega}_1^{\prime }=\vec{\omega}_1-{\frac{d_1}{(2I_1)}}\vec{n}\times
\Delta \vec{P}~,  \label{equationb}
\end{equation}
\begin{equation}
\vec{v}_2^{\prime }=\vec{v}_2-\Delta \vec{P}/m_2~,~{\rm and} \label{equationc}
\end{equation}
\begin{equation}
\vec{\omega}_2^{\prime }=\vec{\omega}_2-{\frac{d_2}{(2I_2)}}\vec{n}\times
\Delta \vec{P}~,  \label{equationd}
\end{equation}
where $\vec{v}_i^{\prime }$ and $\vec{\omega}_i^{\prime }$ are the unknown
velocities of particle $i$ after collision. $I_i$ is the moment of
inertia about the center of particle $i$ and $\Delta \vec{P}$ is the change
of linear momentum of particle 1 and is a function of $r $, $\mu $,
and $\beta _0$: 
\begin{equation}  \label{eqP}
\Delta \vec{P}=-m_{\rm 12}(1+r )\vec{v}_c^{(n)}-{\frac 27}m_{\rm 12}(1+\beta )%
\vec{v}_c^{(t)},  \label{eq:DPbeta}
\end{equation}
with the reduced mass $m_{\rm 12}=m_1m_2/(m_1+m_2)$. $(n)$ and $(t)$ indicate 
the normal and the tangential components of $\vec{v}_c$, respectively, and the
factor 2/7 in the tangential part of Eq.\ (\ref{eqP}) stems from the fact
that solid spheres are used \cite{luding98b}. $r$ is the coefficient of
normal restitution and $\beta = \min[\beta_0,\beta_1]$ is the coefficient
of tangential restitution. The latter is simplified so that either sticking 
or sliding are exclusively allowed. A sticking contact has a constant maximum 
tangential restitution $\beta = \beta_0$, with $-1 \le \beta_0 \le 1$ due 
to the elasticity of the material. Typical values for, e.g.~acetate or glass, are 
$\beta_0 \approx 0.5$ \cite{foerster94}. Sliding, Coulomb-type interactions have 
$\beta = \beta_1$, i.e.~$\Delta P^{(t)}$ is limited by $\mu_w \Delta P^{(n)}$. 
Using the basic conservation laws one can find
$\beta_1 = - 1 -\mu (1+r) \cot(\gamma) (1+1/q_i)$, with the collision
angle $\gamma$, and the factor $q_i=4I_i/(m_i d_i^2)$ that accounts for the 
mass distribution inside the particles \cite{foerster94,luding95b,luding98b}.
As illustration, a schematic picture of two 
colliding particles is given in Fig.\ \ref{fig:coll_schem}.
The angular velocities are $\omega_1 = \omega_2 = 0$ immediately before 
collision (a) and non-zero after collision (b).
For a more detailed discussion of the above equations see Ref.\ \cite
{luding95b,luding98b}.

For the simulation of rigid particles, we use an event driven method such 
that the particles undergo an undisturbed motion until an event occurs. 
An event is either the collision of two particles or the
collision of one particle with a wall. From the velocities just before
contact, the particle velocities after a contact are computed following 
Eqs.\ (\ref{equationa})-(\ref{equationd}). Lubachevsky\ \cite{lubachevsky91}
introduced an efficient scalar ED algorithm which updates only 
those particles involved in the previous collision. Like in Refs.\
\cite{luding95b,luding94c} we implement the algorithm of Ref.\ \cite
{lubachevsky91} with some changes and extensions. Despite gravitational
acceleration, all times of contact of particles with each other or with the
lateral walls can be calculated analytically. The
coefficient of normal restitution depends on the partner of the colliding
particle, i.e.~$r$ or $r_w$ is used to indicate particle-particle, or
particle-wall collisions, respectively.

\subsection{The connection between hard- and soft-sphere models}
\label{sec:TCED}

In the ED method, the time during which two particles are in
contact $t_c$ is implicitly zero. The consequence is that exclusively pair
contacts occur and the instantaneous momentum change $\Delta \vec P$ 
in Eqs.\ (\ref{equationa})-(\ref{equationd}) suffices to describe the
collision completely. ED algorithms with constant $r$ run into difficulties 
when the time between events, $t_{\rm ev}$, becomes too small --- typically
in systems with strong dissipation --- and the so-called 
`inelastic collapse' occurs \cite{bernu90,mcnamara92,mcnamara93,luding94}. 
To handle this problem, several attempts have been proposed recently
\cite{luding94,du95,luding96e,giese96,deltour97,grossman97,shattuck97,luding97c,aspelmeier98,luding98f}, 
and one of them \cite{luding96e,luding97c,luding98f} 
switches off dissipation when the collision frequency becomes too large.

In MD simulations of dynamical systems, on the other hand, $t_c > 0$ and 
only a limited amount of energy can be dissipated per contact duration. 
A finite $t_c$ thus implies 
a finite energy dissipation rate. In a dense system of soft particles, 
energy dissipation becomes ineffective, i.e.~the `detachment effect' occurs
\cite{luding94d,luding95}. This effect is not obtained with hard particles
and a constant coefficient of restitution $r$, however, the effect can be
observed when using 
\begin{equation}
r^{(i)}_n = {\left \{
             { \begin{array}{lll}
                 {r} & {\rm ~~for~ } & t^{(i)}_n > t_c\\ 
                 {1} & {\rm ~~for~ } & t^{(i)}_n \le t_c~,
               \end{array} }
           \right . } 
\label{epsnew}
\end{equation}
as the restitution coefficient for the collision $n$ of particle $i$.
In Eq.\ (\ref{epsnew}) $t^{(i)}_n$ is the time since the last collision
and $t_c$ is the threshold for elastic contacts that can be identified
(up to a constant factor of order unity) with the contact duration in 
the soft particle model. Thus, an additional material parameter is
defined for the hard sphere model, that leads to qualitative agreement
between ED and MD simulations and in addition avoids the inelastic collapse.
Note that $t_c$ has different physical meaning in either hard or soft 
sphere model.
The traditional ED method has $t_c = 0$. The extended model that uses
Eq.\ (\ref{epsnew}) with $t_c > 0$ is in the following referred to as the 
TCED method.

The integral of all forces $\vec f(t)$, acting on a particle at time 
$t \in [t_0,~t_c]$, is needed to calculate the momentum change of this particle
in the framework of the soft particle model:
\begin{equation}
\Delta \vec P = \int_{\rm t_0}^{t_0+t_c} \vec f(t) {\rm d}t~.
\end{equation}
In general, the contact begins at time $t_0$ and ends at time $t_0+t_c$.
For a constant force $\vec f$ or an infinitesimally small time
interval $t_c$, the momentum change $\Delta \vec P$ in 
Eqs.\ (\ref{equationa})-(\ref{equationd}) can be replaced by the
term $f(t) $d$t$ to arrive at a differential formulation for
the change of velocities $\vec v' - \vec v$ and $\vec \omega' - 
\vec \omega$. The primed and unprimed quantities are the values
at time $t_0+t_c$ and $t_0$, respectively.

The stress tensor, defined for a test volume $V_c$, can be
written as
\begin{equation}
\sigma_{\alpha \beta} = \frac{1}{V_c}
   \left [
      \sum_j r_\alpha f_\beta - \sum_i m_i v_\alpha v_\beta
   \right ] .
\label{eq:sigma}
\end{equation}
The indices $\alpha$ and $\beta$ are the Cartesian coordinates, the
$r_\alpha$ are the components of the vector from the center of mass
of a particle to the point $j$, where a force with components
$f_\beta$ acts. Particle $i$ has the mass $m_i$ and a velocity
with the components $v_\alpha$. The first sum runs over all contact
points $j$, and the second sum runs over all particles $i$, both
within $V_c$.
In the static limit, the second term vanishes, since all velocities
vanish. On the other hand, for a hard-sphere gas, the left term
has to be treated differently, since no forces are defined. The
dynamic equivalent to $f_\beta$ is the change of momentum per
unit time $\Delta P_\beta / \Delta t$. For a hard-sphere gas
the stress due to collisions may be evaluated as the average over
all collisions in the time interval $\Delta t$. 
The first sum runs over all collisions taking place in the
time between $t-\Delta t$ and $t$.
In general, the volume $V_c$ and the time-interval
$\Delta t$ have to be chosen large enough to allow averages
over enough particles and enough collisions.

\subsection{The time driven, soft particle technique}

Even without using the soft particle method in this study, it is 
convenient to discuss briefly the standard interaction forces
and their connection to the hard sphere collision operator
that involves the total momentum change $\Delta \vec P$.
Replacing $\Delta \vec P$ in Eqs.\ (\ref{equationa})-(\ref{equationd})
by $\vec f(t) t_{\rm MD}$, with the molecular dynamics time step
$t_{\rm MD}$, allows the integration of the corresponding equations
of motion with standard numerical methods \cite{allen87,rapaport95}.

Since the modeling of realistic deformations of the particles would be
much too complicated, let us assume that the overlap of two particles
is the quantity important for the interaction potential. 
The interaction is short range, i.e.~the particles
interact only when they are in contact so that their penetration depth 
$\delta = {1 \over 2}(d_1 + d_2) - (\vec r_1 - \vec r_2) \cdot \vec n$ is positive.

The first force, acting from particle 2 on particle 1 --- accounting for the excluded 
volume which each particle occupies --- is an elastic repulsive force
\begin{equation}
\vec f_{\rm el} = k_n \delta_0 (\delta/\delta_0)^{\nu} \vec n~,
\label{eq:fn}
\end{equation}
where $k_n$ is the elastic modulus and $\delta_0$ is a normalization 
constant dependent on the nonlinearity $\nu$ and the dimension.
In the simplest case of a linear spring that follows Hooke's law,
$\nu = 1$, in the case of elastic spheres in three dimensions,
$\nu = 3/2$, i.e.~a Hertz contact \cite{landau75}, and for conical contacts,
$\nu = 2$ can be used.

The second force --- accounting for dissipation in the normal direction
--- is a viscous damping force
\begin{equation}
\vec f_{\rm diss} = \gamma_n \dot \delta (\delta/\delta_0)^{\phi} 
  \vec n~,
\label{eq:fdiss}
\end{equation}
where $\gamma_n$ is a phenomenological viscous dissipation coefficient and 
$\dot \delta = - \vec v_{12} \cdot \vec n = - (\vec v_1 - \vec v_2) \cdot \vec n$ 
is the relative velocity in the normal direction.

The simple linear spring-dashpot model (with $\nu=1$ and $\phi=0$)
can be solved analytically and leads to 
a contact duration $t_c = \pi/\omega$ and a restitution 
coefficient $r = \exp(-\pi \eta / \omega)$, with $\omega = 
\sqrt{\omega_0^2-\eta^2}$, $\omega_0^2 = k_n / m_{12}$, $\eta = \gamma_n
/ (2 m_{12})$, and $m_{12} = m_1 m_2 / (m_1 + m_2)$ \cite{luding98b}.
A nonlinear repulsive potential can at least be solved in the limit
$\gamma_n \rightarrow 0$ and the dependency of $r$ on the velocity
of impact can be estimated with reasonable accuracy 
\cite{kuwabara87,luding94d,brilliantov96}.

The third force --- accounting for friction in a simplified way
--- acts in the tangential direction and can be chosen in the simplest 
case as
\begin{equation}
\vec f_{\rm shear} = - \gamma_t \dot \xi \vec t~,
\label{eq:ftvisc}
\end{equation}
where $\gamma_t$ is the viscous damping coefficient in 
tangential direction and $\dot \xi = \vec v_{12} \cdot \hat t$ is the 
tangential component of the relative velocity, with 
$\hat t = \vec v_{12} / |\vec v_{12}|$.
Eq.\ (\ref{eq:ftvisc}) is a rather simplistic description of shear 
friction. For many applications (arching, heap formation)
it is, however, important to include more realistic static
friction \cite{lee93,wolf97} what can be realized by a virtual tangential
spring \cite{cundall79,schafer96}:
when two particles start to touch each other, one puts a virtual
spring between the contact points of the two particles, and
$\vec \xi(t_1) = \int_{t_0}^{t_1} [\vec v_{12}(t) \cdot \hat t(t)] {\rm d}t ~ \hat t(t_1)$ 
is the {\it total tangential displacement} of this spring during the contact. 
The restoring frictional force is thus $- k_t \vec \xi$.
According to Coulomb's criterion, the maximum value of the restoring 
force is proportional to the normal force $f^c_n$ at this contact,
with the friction coefficient $\mu$. Cast into a formula this
gives a friction force
\begin{equation}
\vec f_{\rm fric} = - \frac{\vec \xi}{|\vec \xi|} 
                                \min(k_t |\vec \xi|, \mu f^c_n)~.
\label{eq:ftfric}
\end{equation}
Note that the tangential spring has to be kept at a maximum length
$\xi_{\rm max} = \mu f^c_n / k_t$ in order to lead to reasonable agreement
with contact dynamics simulations or theoretical calculations \cite{radjai97c}.
Only when particles are no longer in contact with each other is the spring removed.
The main source of static friction in real systems is the geometrical roughness of 
the surfaces \cite{poschel93,walton93c,walton94,poschel95},
and the same effects of particle stopping can be
obtained also without Eq.\ (\ref{eq:ftfric}) by using particles of
complicated shapes, like crosses or polygons 
\cite{buchholtz94,buchholtz95,kohring95,matuttis97}.
In fact, when particles deviate from the spherical shape,
rotations are suppressed in dense packings under strong load.
However, in some cases it is sufficient to use
a combination of Eqs.\ (\ref{eq:ftvisc}) and (\ref{eq:ftfric}):
\begin{equation}
\vec f_{\rm dyn} = - \min(\gamma_t \dot \xi, \mu f^c_n) \vec t~,
\label{eq:ftdyn}
\end{equation}
a rather bold alternative to the more realistic static
friction law in Eq.\ (\ref{eq:ftfric}), but sufficient for many, 
especially dynamic situations \cite{radjai97c}. For a comparison 
of ED and MD simulation results, see Refs.\ 
\cite{luding94c,luding94d,radjai97c,brendel98}.

\subsection{DSMC simulation method}
\label{sec:DSMC}

Direct simulation Monte Carlo (DSMC) was first proposed 
for the simulation of rarefied gas flows \cite{bird94}
and is also used for liquid-solid flow simulations, see
Ref. \cite{tanaka96} and references therein. One of the 
algorithm's advantages is its suitability for parallelization,
what makes it a convenient tool for the modeling of granular media
\cite{muller97,luding98b}. 

In DSMC the evolution of the system is integrated in time steps 
$\tau$, at each of which every particle is first moved without
feeling other particles. The particles are then sorted into
square cells with sides $L_c$ and volume $V_c=L_c^2$.
$L_c$ is set as one half of the mean free path, but never smaller
than two bead diameters. The time step $\tau$ is always chosen small
enough to assure that even the fastest particle needs several
time steps to cross a cell.
Between particles in the same cell stochastic collisions take place; the
rules for these collisions are taken from kinetic theory. First we choose 
the number of collision pairs in each cell,
\begin{equation}\label{EQ:MC}
  M_{c} = \frac{N^c (N^c-1)\sigma v_{max}\tau}{2V_c}~,
\end{equation}
where $N^c$ is the number of particles in the
cell, $v_{max}$ is an upper limit for the relative velocity between
the particles, and $\sigma=2d$ is the scattering cross section of 
discs in 2D. To get $v_{max}$ we sample the velocity distribution from time
to time and set $v_{max}$ to be twice the maximum particle velocity
found. In order to account for the actual relative
velocities we apply an acceptance-rejection method: for a pair of
particles $i$ and $j$ the collision is performed if
\begin{equation}
  \frac{|\vec{v}_i-\vec{v}_j|}{v_{max}} < Z~,
\end{equation}
where $Z$ is uniformly distributed in the interval $[0.0, 1.0[$. 
This method leads to a collision probability proportional to the relative
velocity of the particles. 

Since the collision takes place regardless of the
position in the cell, we have to choose an impact
parameter $b$ in order to calculate the post-collision
velocities.  The impact parameter is defined as
\begin{equation}
b = \left | ( {\vec r}_i - {\vec r}_j ) \times 
      \frac{( {\vec v}_i - {\vec v}_j )}{| {\vec v}_i - {\vec v}_j |}
    \right |
  = d \sin \gamma~,
\end{equation}
where $\gamma$ is the angle between $( {\vec v}_i - {\vec v}_j )$
and $({\vec r}_i - {\vec r}_j)$.  For central collisions $b=0$, and
$b=d$ for grazing collisions. Assuming molecular chaos, $b$ is drawn from a
uniform distribution in the interval $[-d, d]$.  
The rest of the collision scheme is identical to the event driven
procedure, so  that the normal component of the post-collision velocity
is ${\vec{v'}}^{(n)}=-r \vec v^{(n)}$, whereas the tangential
component remains unchanged, i.e.~$\beta_0 = -1$. 

To get better results at higher densities the DSMC method was
modified in two respects. First, the number of collisions $M_c$ in
equation (\ref{EQ:MC}) is increased by replacing the volume $V_c$ of 
a cell with the effective free volume $V_c-V_{0}$, where $V_{0}$ is 
the volume the particles in that cell would need in a
random close packing  with packing fraction 0.82 in 
2D \cite{gervois92}. Second, we
added an offset of $d$ to the particle distance along the direction of the
momentum transfer after the collision \cite{alexander95b}. 

\section{Comparison of ED and DSMC simulations}
\label{sec:EDvsDSMC}

In this section two simulations are presented, starting with
the same initial condition, and using the same parameters, but being
carried out with the ED and the DSMC methods. In ED the probability 
distribution of the impact parameter may deviate from the case 
expected for molecular chaos, whereas DSMC always uses 
a constant probability for $b$ in 2D. The simulations involve
$N = 99856=316^2$ dissipative particles with restitution coefficient 
$r=0.8$ in a periodic quadratic system with volume fraction
$\varrho = 0.25$. The system size is $l = L d$ with dimensionless
size $L$ and particle diameter $d$. In order to reach
an equilibrated initial condition, the system is first
allowed to evolve with $r=1$ for about 10 collisions per 
particle, so that a Maxwellian velocity distribution and
a rather homogeneous density distribution exists. Then,
at $t=0$\,s, dissipation is set to $r=0.8$ and the quantities
of interest are calculated as functions  of time.

\subsection{Freely cooling granular materials}

In the homogeneous cooling state
\cite{haff83,goldhirsch93,goldhirsch93b,mcnamara96,luding98c}
we expect that the energy $K(t)$ of the system decays with time
and follows the functional form
\begin{equation}
\frac{K(t)}{K_0} = \left ( \frac{1}
                            {1+t/t_0}
                    \right )^2 ~~,
\label{eq:kt}
\end{equation}
with the theoretically expected time scale 
\begin{equation}
t_0 = \frac{\sqrt{\pi} d s_{*}(\varrho)}{2 (1-r^2) \varrho \bar v}
\end{equation}
as a function of the initial mean velocity $\bar v = \sqrt{2 K_0/ N m~}$, 
the initial kinetic energy $K_0 = K(0)$, 
the particle diameter $d$, the restitution coefficient $r$, and the 
volume fraction $\varrho$ with $s_{*}(\varrho) = (1-\varrho)^2/(1-7 \varrho/16)$
\cite{haff83,jenkins85b,mcnamara96,luding98d}. 
Inserting the corresponding parameters $1-r^2=0.36$, 
$s_{*}(\varrho) \approx 0.63158$, $d=0.001$\,m, and $\bar v =0.2047$\,m/s,
we have $t_0^{-1} = 23.24$\,s$^{-1}$.

\begin{figure}[t]
\begin{center}
\epsfig{file=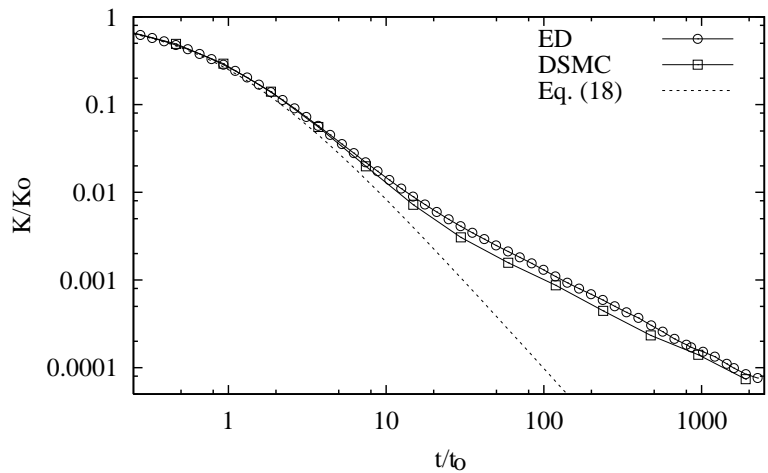,height=11.0cm,clip=,angle=-90}
\end{center}
\caption{\label{fig:kinenergy} Normalized kinetic energy
  vs. normalized time from an ED and a DSMC simulation in 2D with 
identical initial conditions and $N = 99856$, $\varrho = 0.25$, 
$r=0.8$. The dashed line represents Eq.\ (\protect{\ref{eq:kt}}).
The 
}
\label{fig:energy}
\end{figure}

In Fig.\ \ref{fig:energy} we present the normalized
kinetic energy $K(t)/K_0$ as a function of the normalized
time $t/t_0$.
At the beginning of the simulation we observe a perfect agreement
between the theory for homogeneous cooling and both simulations.
At $t/t_0 \approx 2$ substantial deviations from the homogeneous 
cooling behavior become evident, and only at $t/t_0 \approx 10$
a difference between ED and DSMC can be observed. After that time,
the kinetic energy obtained from the DSMC simulation is 
systematically smaller than $K(t)$ from the ED simulation.
We relate this to the fact that the molecular chaos assumption 
of a constant probability distribution of the impact parameter 
$b$ is no longer valid. Since dissipation acts only at
the normal component of the relative velocity, DSMC dissipates more energy
than ED as soon as the number of central collisions is overestimated
\cite{luding98}. To verify this assumption we take a closer look at the 
impact parameter and its probability distribution in the next subsection.

\subsection{The impact parameter}
\label{sec:collpara}
One basic assumption connected to molecular chaos is
a uniform probability distribution of the impact parameter.
We define $P(b/d)$ to be the probability 
distribution of $b$ and normalize it
such that $\int_0^1 {\rm d}x P(x) = 1$. 
From ED simulations with 
elastic particles the normalized probability distributions 
$P(b/d) = 1$ in 2D, and $P(b/d)=2b/d$ in 3D, are found as 
expected for the case of molecular chaos. 

\begin{figure}[tb]
\begin{center}
\epsfig{file=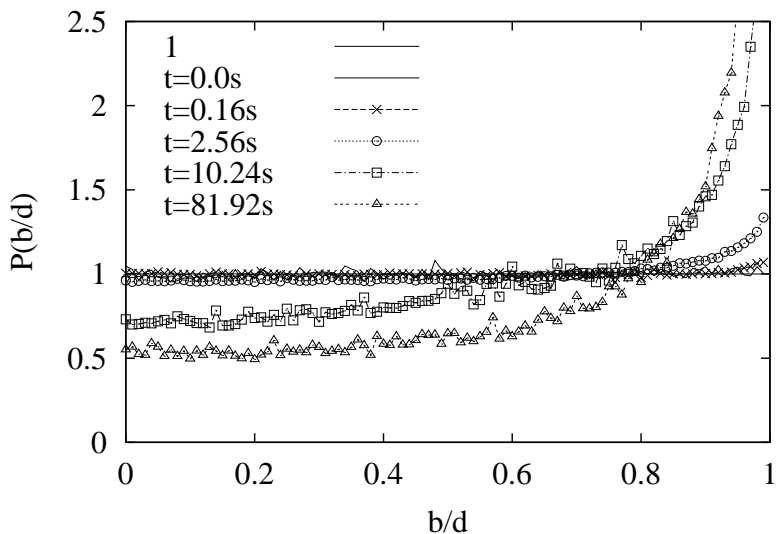,height=11.0cm,clip=,angle=-90}
\end{center}
\vspace{-1.1cm}~\\
\caption{Normalized probability distribution of the contact parameter
from an ED simulation in 2D with $N = 99856$, $\varrho = 0.25$, and $r=0.8$
at different times, as given in the inset.}
\label{fig:pbd1}
\end{figure}

The ED simulation of Fig.\ \ref{fig:kinenergy} leads to
$P(b/d) = 1$ for short times only. For larger times we observe
an increasing (decreasing) probability of grazing (central) collisions.
In Fig.\ \ref{fig:pbd1}, data of the probability distribution 
are presented at different times during the simulation of Fig.\ 
\ref{fig:kinenergy}. As it is obvious from the data, more and more grazing 
collisions occur with increasing simulation time. Evidently,
the assumption of a constant probability distribution of
the impact parameter is violated. 

%\subsection{The reason for the breakdown of molecular chaos}

\begin{figure}[hbt]
\begin{center}
\epsfig{file=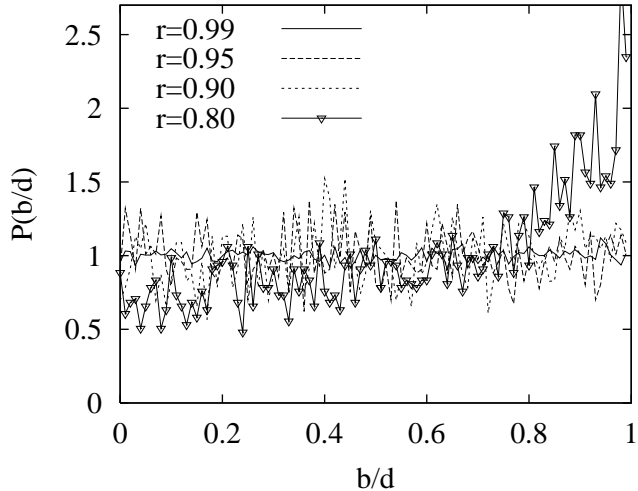,height=9.1cm,clip=,angle=-90}
\end{center}
\vspace{-1.1cm}~\\
\caption{Normalized probability distribution $P(b/d)$ from ED simulations
in 2D with $N = 240$, $\varrho = 0.7495$ and different $r$.
}
\label{fig:pbd2}
\end{figure}

One can imagine at least two reasons for the deviation of 
$P(b/d)$ from the constant value. The first is that $P(b/d)$
might be a function of the density, and that due to density fluctuations,
the form of $P(b/d)$ changes. Thus we calculate $P(b/d)$ in smaller
systems with $N=240$, $r=1$, and different volume fractions, ranging
from very dilute to extremely dense systems. $P(b/d)$ is not sensitive to 
the density, as long as the collisions are elastic \cite{luding98b}. 
Another reason for $P(b/d)$ to deviate from unity might be
dissipation. 
In Fig.\ \ref{fig:pbd2} the restitution coefficient is varied
for fixed $\varrho = 0.7495$. For weak dissipation, i.e.~$r \ge 0.9$,
the distribution is homogeneous. For stronger dissipation $r=0.80$
we find an increasing probability of grazing contacts.

The assumption $P(b/d)=1$ is true in elastic systems for 
arbitrary density. For inelastic systems, $P(b/d)$ is constant for
sufficiently weak dissipation but depends on $b/d$ for strong
dissipation. The breakdown of molecular chaos is {\em not
due to high density}, and also {\em dissipation is not the only 
reason} for it, since the dissipation must be strong enough to cause the 
inhomogeneous distribution.
The remaining question is: why do we observe this increasing probability
of grazing contacts? 

\begin{figure*}[tbp]
\begin{center}
 \epsfig{file=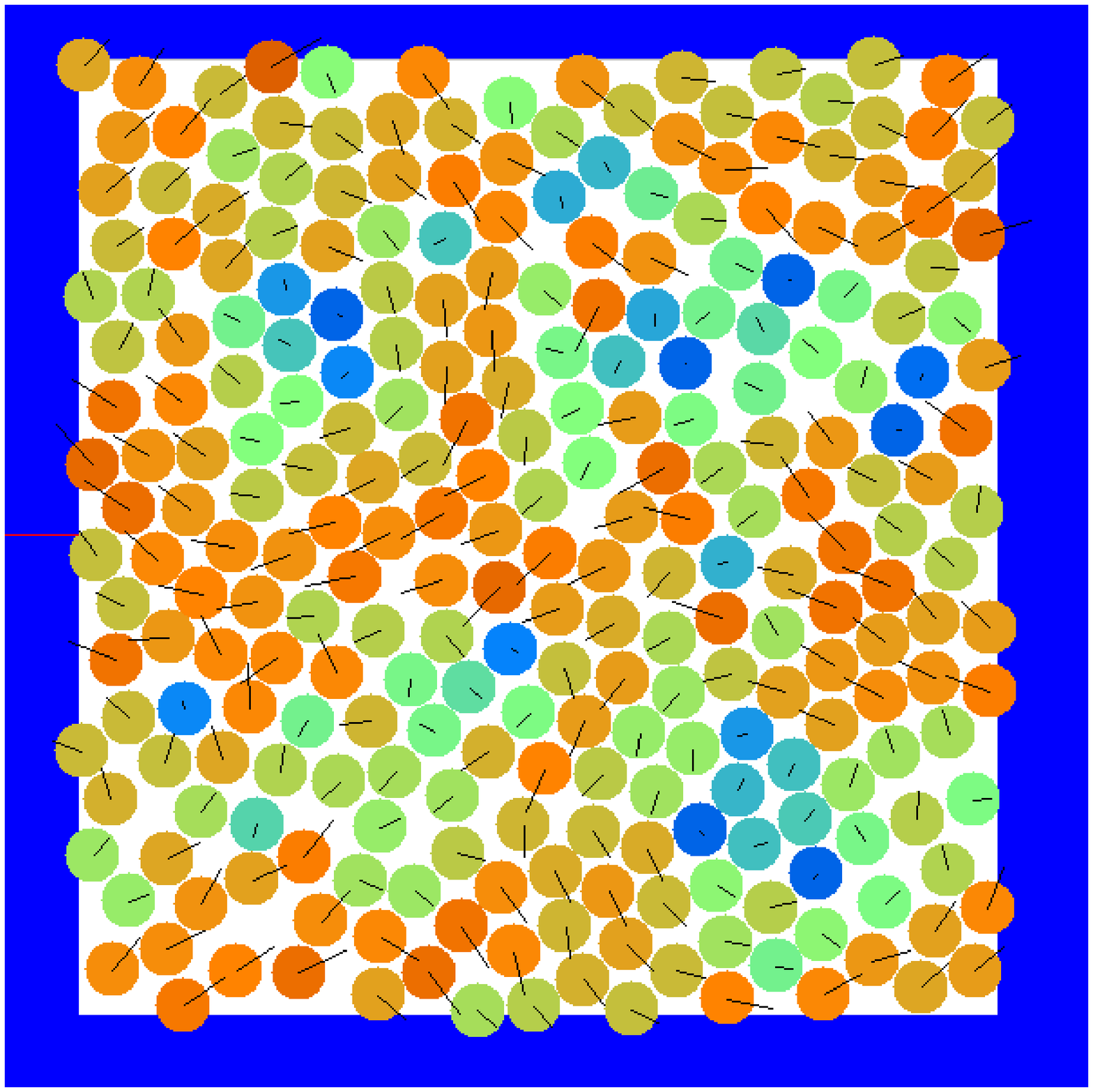,height=5.0cm,clip=,angle=0}
 \epsfig{file=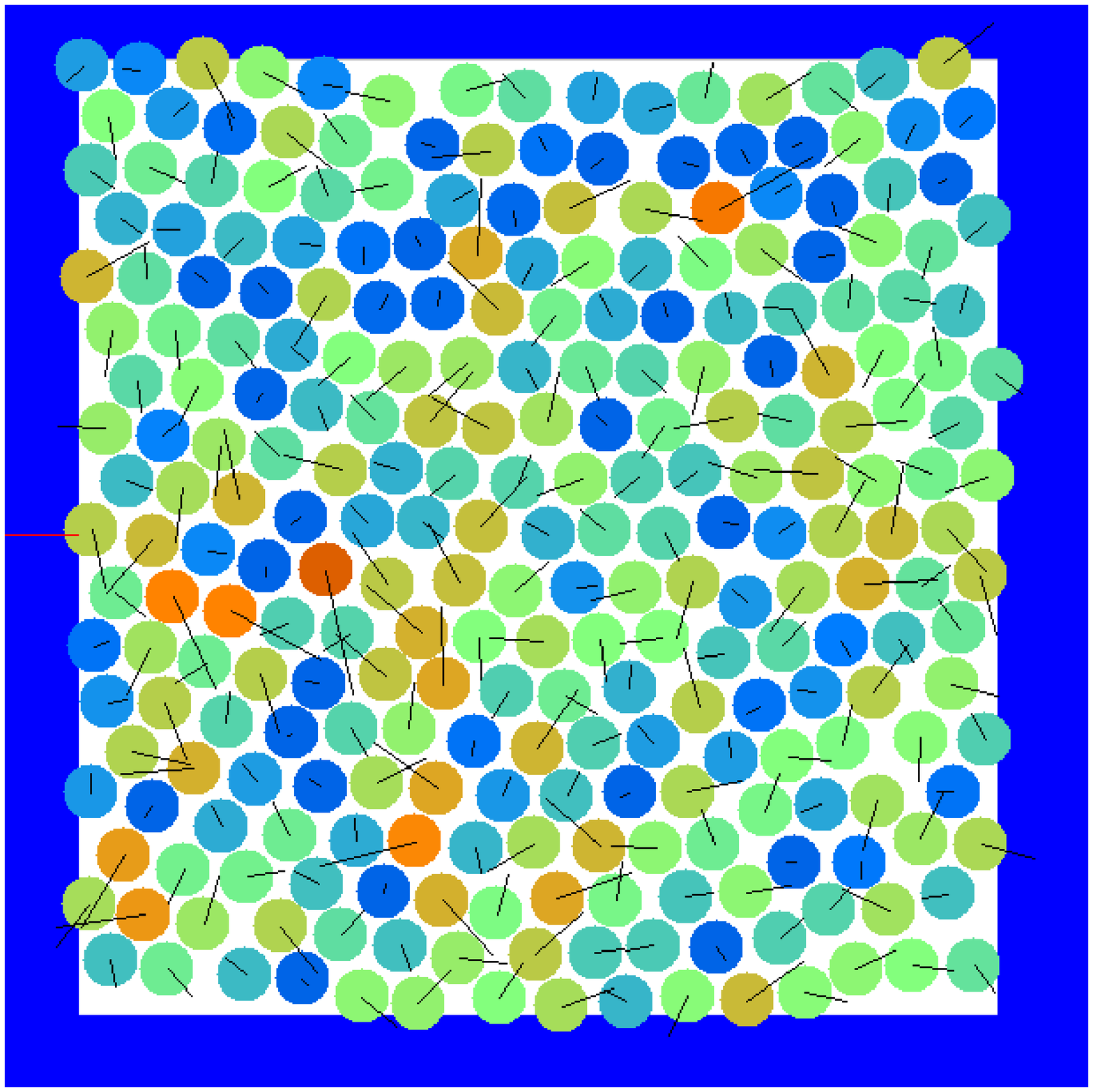,height=5.0cm,clip=,angle=0}
 \epsfig{file=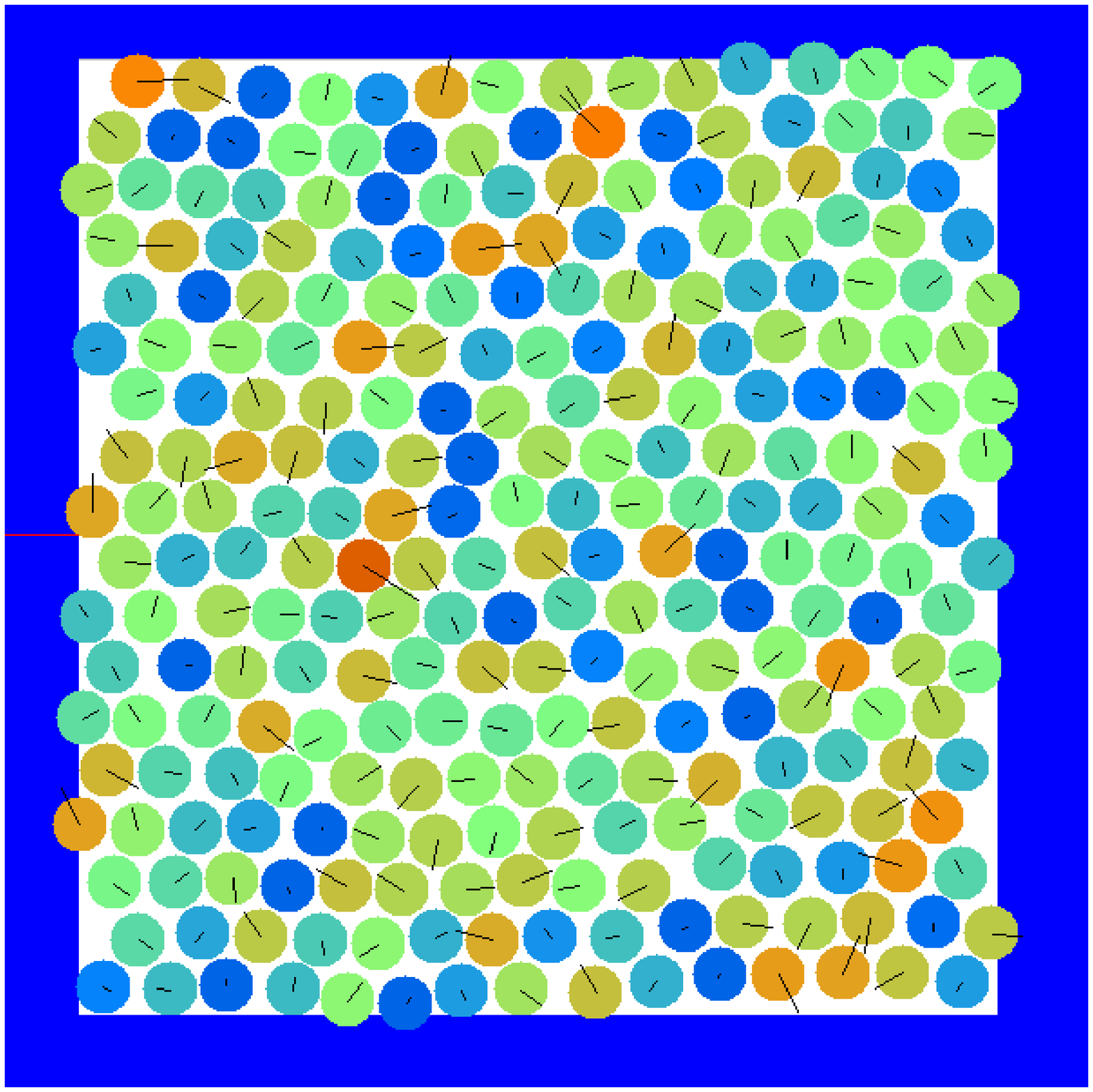,height=5.0cm,clip=,angle=0}
\end{center} 
\vspace{-1.4cm}~\\
\begin{center}
 \epsfig{file=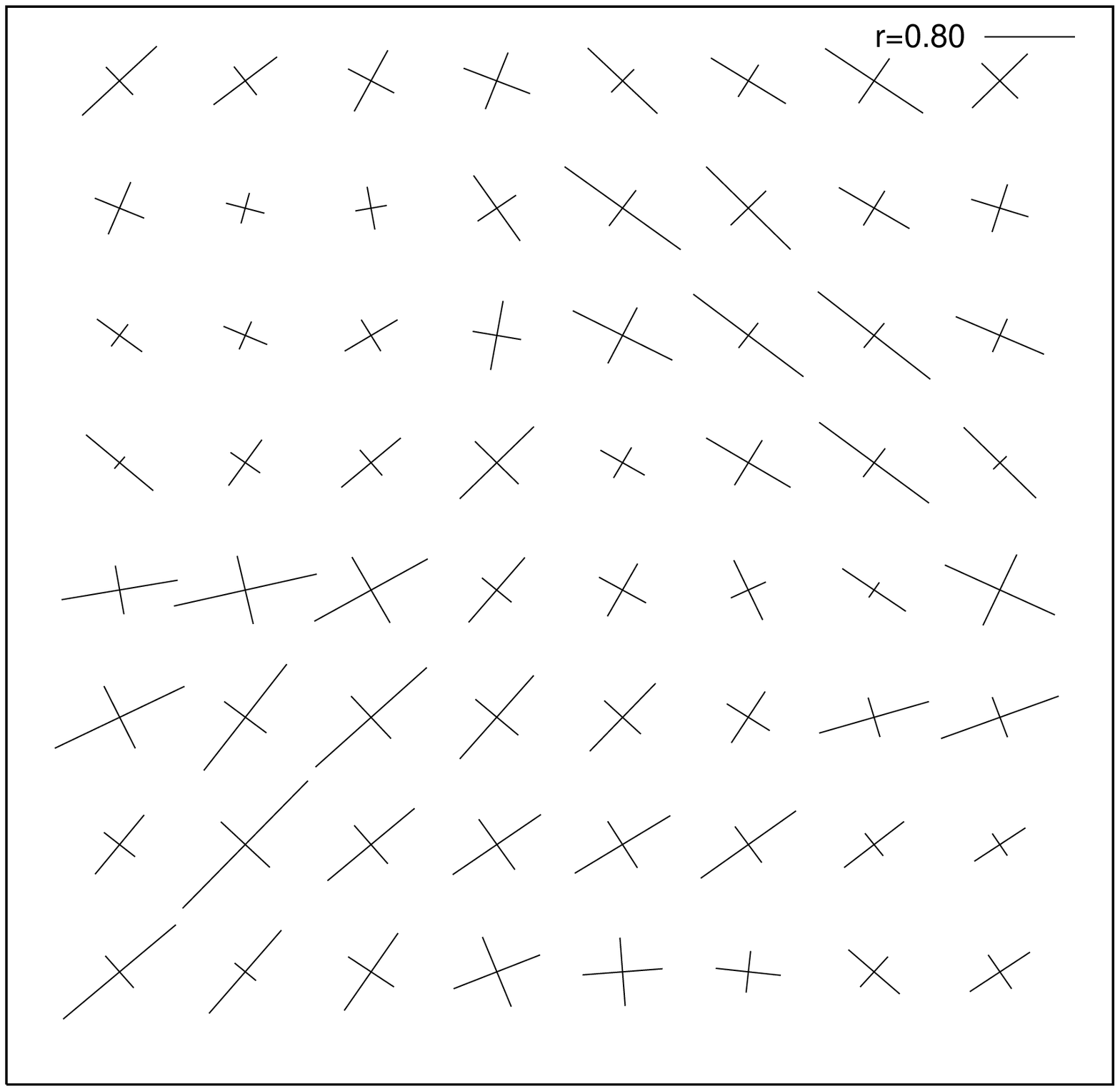,height=4.7cm,clip=,angle=-90} ~
 \epsfig{file=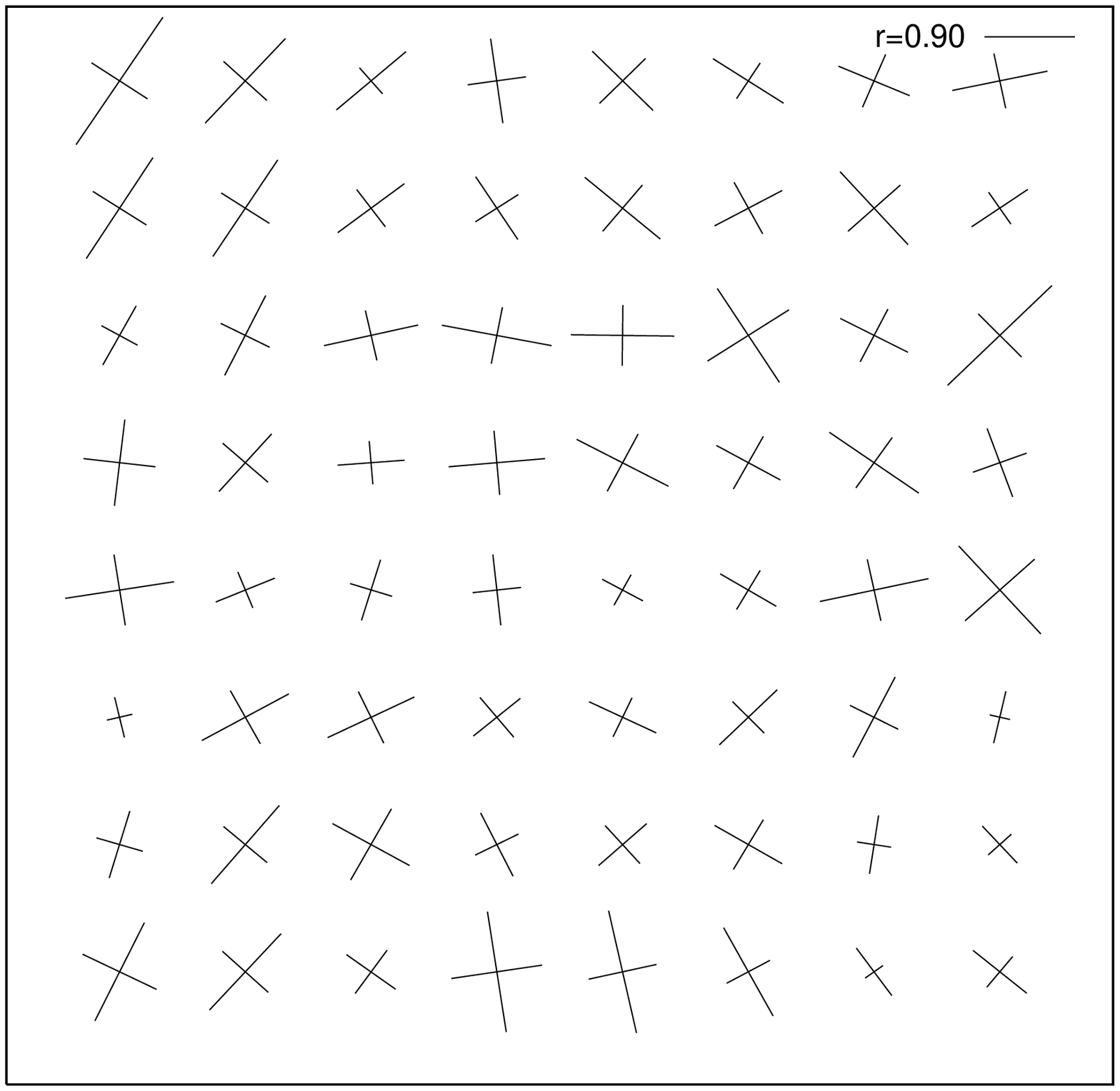,height=4.7cm,clip=,angle=-90} ~
 \epsfig{file=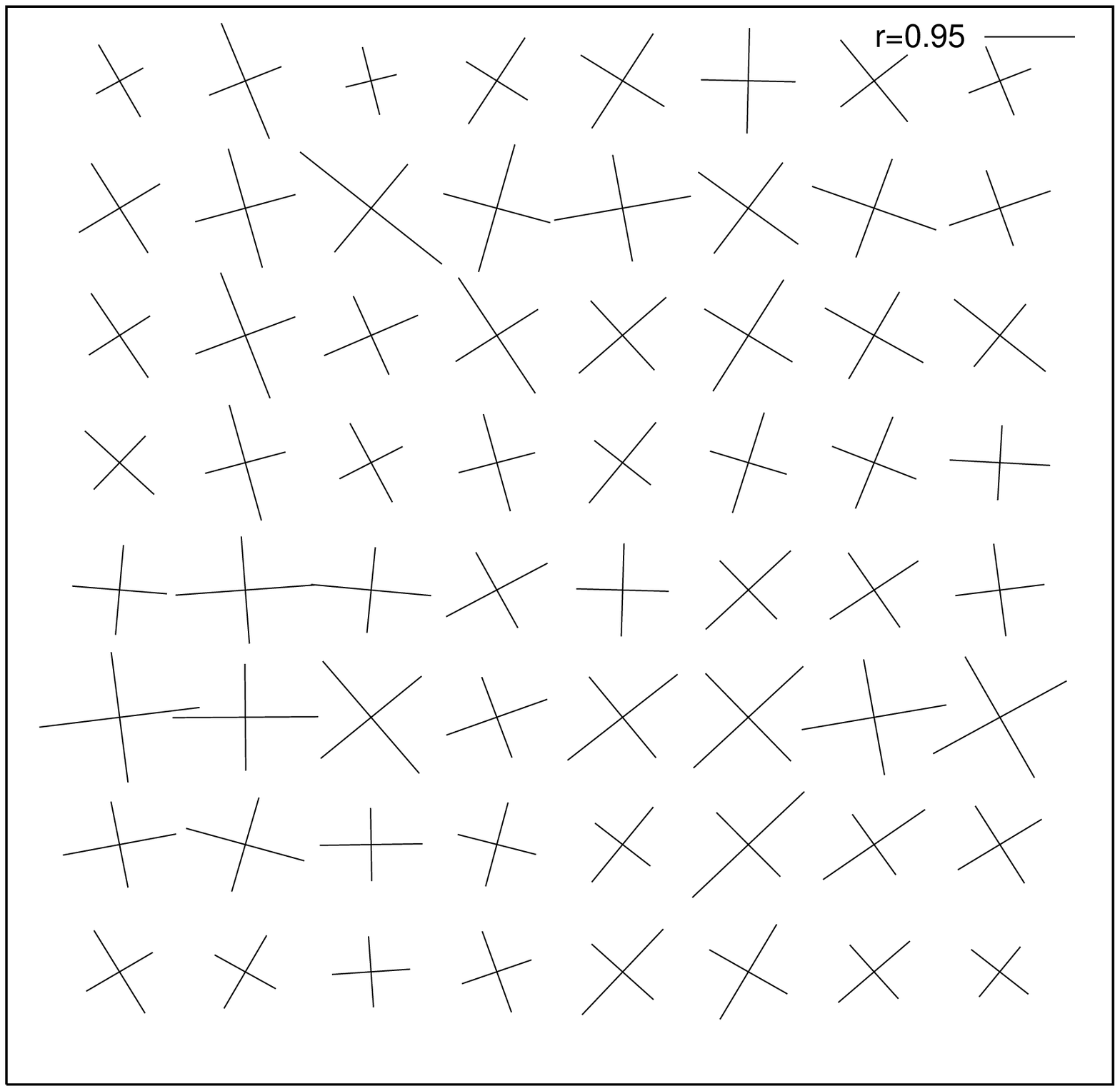,height=4.7cm,clip=,angle=-90}
\end{center}
\vspace{-0.5cm}~\\
\caption{Snapshots from ED simulations in 2D with $N = 240$, 
$\varrho = 0.7495$ and different $r=0.80$ (left), $r=0.90$ (middle), 
$r=0.95$ (right) after $t=1$\,s. The lines give the velocity vector
scaled by 1, 3, and 7 in the left, middle and right panel,
respectively. In the panels below the stress is plotted in the
principal axis representation following Eq.\ (\ref{eq:sigma}).
The average is taken in the time interval $0.5$\,s $\le t \le 1.0$\,s
over all collisions in each cell. The maximum eigenvalues $\sigma_{\rm max}$, 
measured in the left, middle, and right panel are 0.894, 0.568, and 1.56, 
respectively. They are measured in arbitrary units and the figures are 
scaled so that $\sigma_{\rm max}$ has the same length.
}
\label{fig:pbd3}
\end{figure*}

Looking in more detail at the simulations in Fig.\ \ref{fig:pbd2},
we observe that the inhomogeneous distribution for $r=0.8$ is connected
to shear motion of the particles, whereas no visible shear motion
occurs for $r \ge 0.9$. The shear motion in the left panel of
Fig.\ \ref{fig:pbd3} can be understood as the geometrical reason 
for the higher probability for grazing contacts, i.e.~layers of 
particles shear against each other and therefore touch preferentially
with large impact parameter $b$.
The two eddies (top left and bottom right) in the left panel are
rotating in the same sense and lead to small stresses in the interior.
Outside, where the velocities are less correlated one observes larger
stresses and the stress tensor typically has strongly different eigenvalues. 
The more disordered situations in the middle and right panel, in contrast,  
are connected to rather random stresses.

\subsection{The structure factor}

\begin{figure}[htbp]
\begin{center}
\epsfig{file=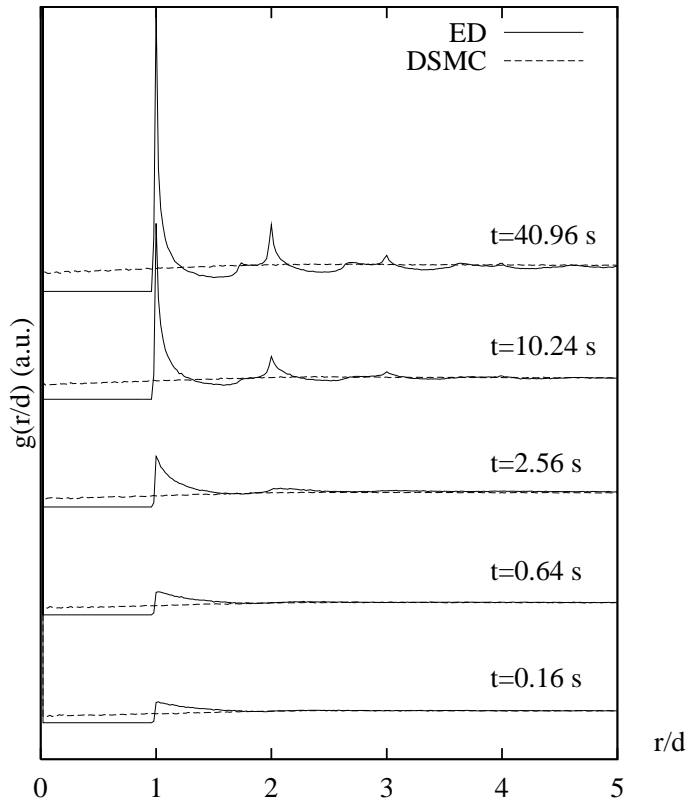,height=8.5cm,angle=-90}
~\vspace{-0.8cm} \\
\end{center}
\caption{Correlation function $g(r/d)$ as obtained from the 
  ED (solid line) and DSMC (dashed line) simulations of 
  Fig.\ \ref{fig:kinenergy}. The different curves are shifted vertically
  in order to avoid overlap. The solid line at $0 < r/d < 1$ gives
  the baseline $g(r/d)=0$.
}
\label{fig:corr}
\end{figure}

One difference between ED and DSMC simulations is the handling
of excluded volume by the two methods. While ED handles
hard spheres with a well defined excluded volume, the DSMC
method models point particles and excluded volume is introduced
by the approximations described in subsection\ \ref{sec:DSMC}.
As expected, we obtain dramatic differences in the particle-particle
correlation function $g(r/d)$ in Fig.\ \ref{fig:corr}, where $r/d$ is
the particle separation in units of particle diameters $d$.
Since ED (solid lines) models hard spheres, $g(r/d)=0$ when $0 \le r/d < 1$.
For short times one observes $g$ as for an elastic hard sphere gas
and at larger times $g(r/d)$ shows a rich structure with peaks at
1, 2, 3, ... and multiples of $\sqrt{3}/2$ (factor $\ge 2$), 
indicating a rather close triangular packing of monodisperse spheres. 
In contrast, the
DSMC simulations (dashed lines) show no short range correlations 
between particle positions throughout the whole simulation.

The next question is, whether this difference has
consequences at greater length scales. The formation and growth of
large clusters \cite{goldhirsch93,goldhirsch93b,mcnamara96}
is quantified by the large wavelength modes of $g(r/d)$, or equivalently,
by the structure factor $S(k)$ at small wave-number $k$. We calculate $S(k)$ 
by a direct FFT (fast Fourier transform) of the two-dimensional density. 
Before we apply the FFT we map the particles onto
a $M \times M$ lattice, where $M$ is the closest power of 2 that gives 
a lattice box size of about one diameter.

\begin{figure*}[t]
\begin{center}
\epsfig{file=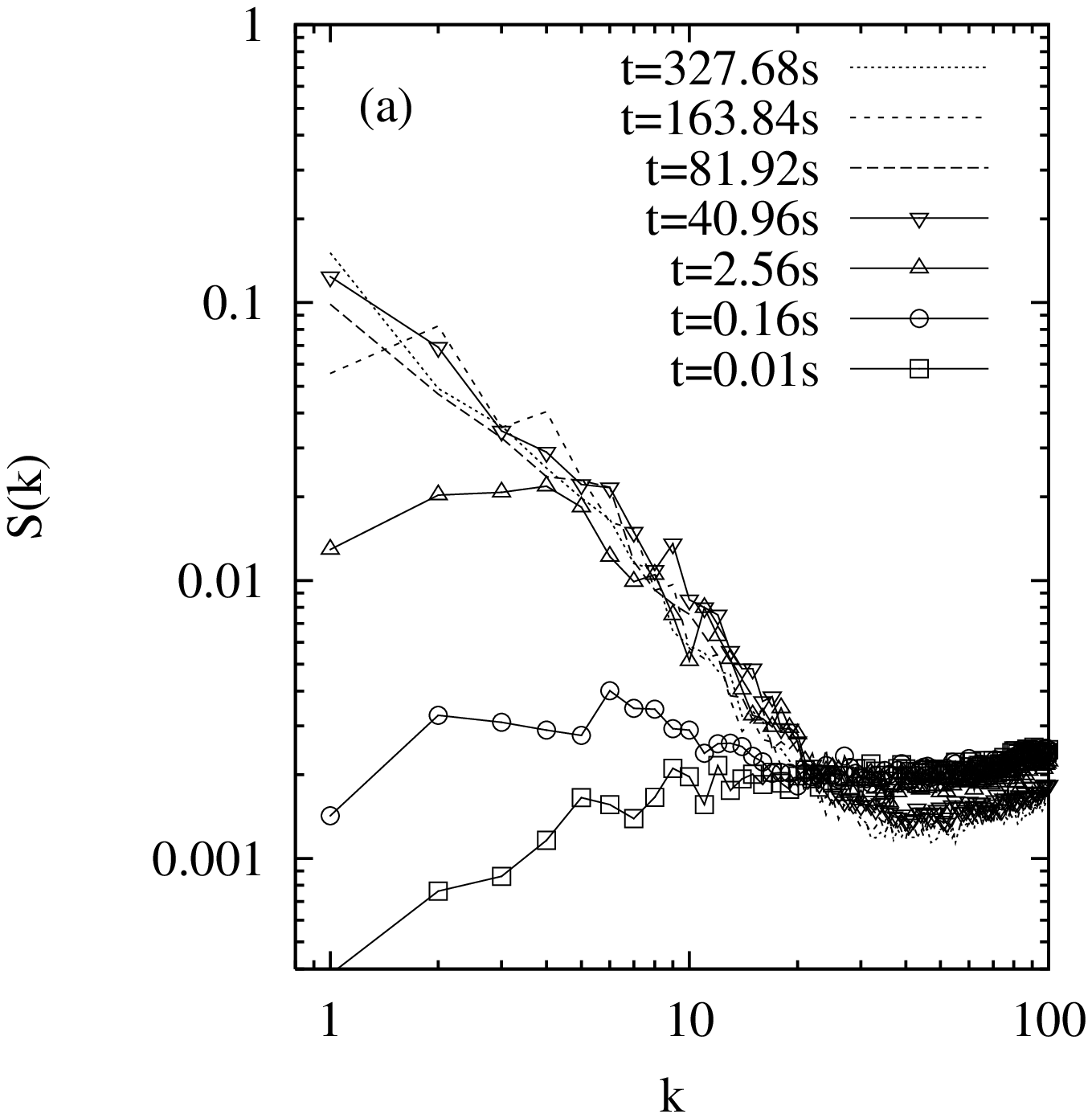,height=7.5cm,angle=-90}
\epsfig{file=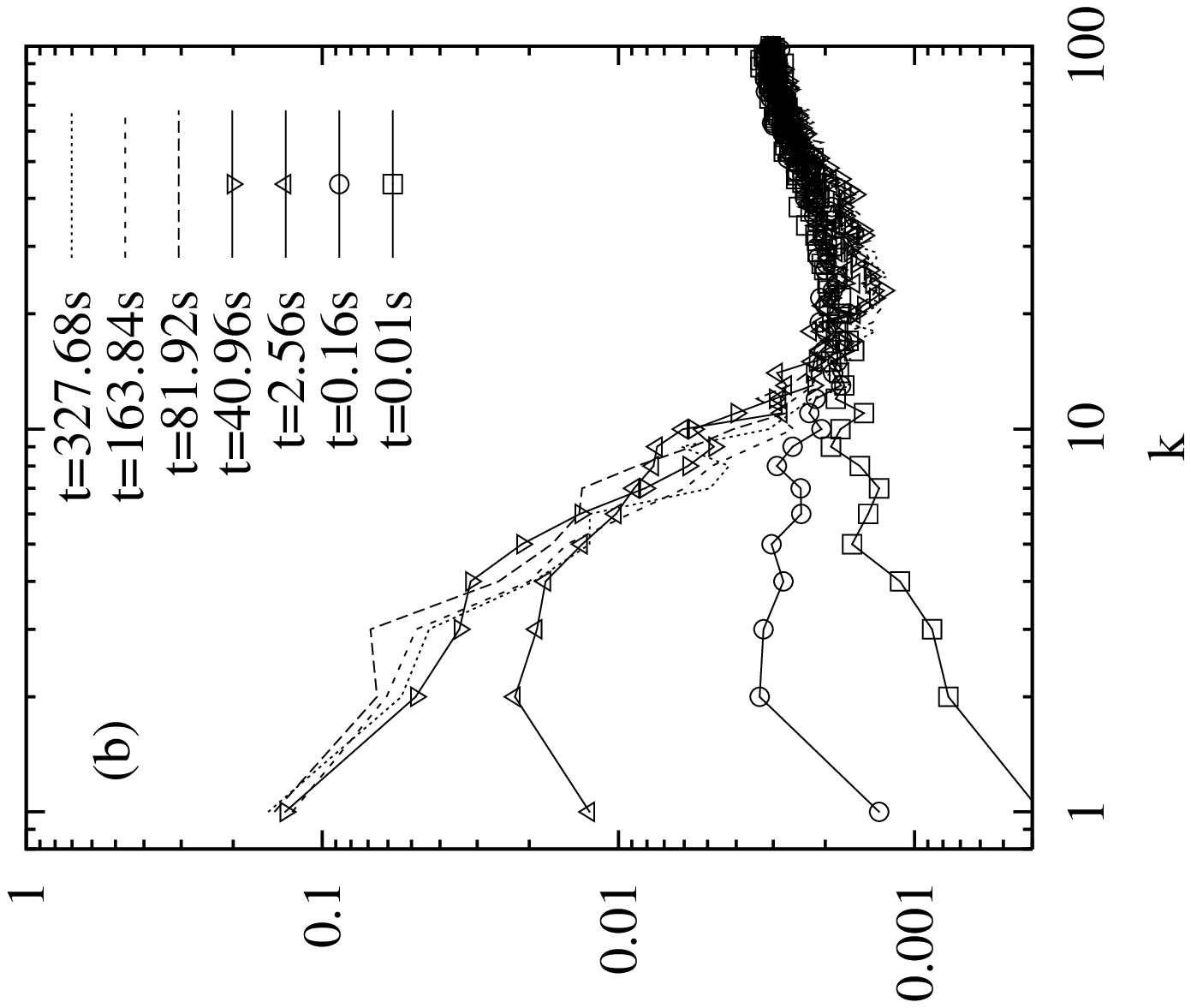,height=7.5cm,angle=-90}
\end{center}
\caption{(a) Structure factor obtained from the ED simulations of
  Fig. \ref{fig:kinenergy} as function of the wavenumber
  $k=l/\lambda$, with wavelength $\lambda$ and system size $l$.
(b) Structure factor obtained from the corresponding DSMC simulation.
}
\label{fig:struct}
\end{figure*}

The structure factors obtained by ED are presented in Fig.\ \ref{fig:struct}(a) 
and those obtained by DSMC in Fig.\ \ref{fig:struct}(b). Different symbols 
correspond to different times. We observe an increase of $S(k)$ for short 
wavenumbers 
$k < 25$, until the structure factor ceases to change for $t \ge 20$\,s.
The structure factor agrees reasonably well for both simulation
methods, and for large enough times it does not change further.
This proves that the DSMC simulation is capable to reproduce
the more realistic but computationally more expensive 
ED results that account for the excluded volume by construction.
Even without short-range correlations, the information about
large wavelengths is well reproduced by DSMC simulations.

\section{The clustering instability}
\label{sec:cluster}

The essential difference between a granular and a classical
gas is dissipation. The resulting clustering instability was examined in 1D
\cite{bernu90,mcnamara92,mcnamara93,luding94,grossman96b,kudrolli97,kudrolli97b}
and in 2D 
\cite{sibuya90,goldhirsch93,mcnamara94,trizac95,mcnamara96,spahn97,deltour97,orza97}.
Cluster growth could be described theoretically only in the case of
irreversible aggregation \cite{trizac95,trizac96}, the more general case 
of reversible aggregation is still an open issue.

Detailed examination of the inelastic collapse by McNamara and Young
\cite{mcnamara96} led to the picture of different `phases'. In a periodic 
system without external forcing exists a critical dissipation --- connected to 
volume fraction and restitution coefficient --- above which clustering 
occurs and below which the system stays in molecular chaos.
In the transient regime the system behavior seems to depend on
the system size and  shearing modes are frequently observed.

\subsection{Parameter studies}

In the following we examine periodic systems of length $L = l/d$ in 2D, with the 
particle diameter $d$, using the TCED simulation method as described in
subsection\ \ref{sec:TCED}.
The volume fraction is $\varrho = N \pi (d/2 l)^2 = (\pi/4) N/L^2$
and the particles are initially arranged on a square lattice,
homogeneously distributed in the system. First the system is equilibrated 
with $r=1$, then the dissipation is switched to the desired value
and the simulation starts at $t=0$\,s. Here we use
a rather small system with  $N=784$, $L=50$, and $\varrho \approx 0.25$. 

\begin{figure}[htb]
\vspace{.4cm}~
~~~~~~~~~~~~~~~~~~~~~~~~~~~~~~~~~~~~~~~~~~(a) \hfill
(b)~~~~~~~~~~~~~~~~~~~~~~ \vspace{-1.4cm}\\
 ~\hspace{-1cm}~ \epsfig{file=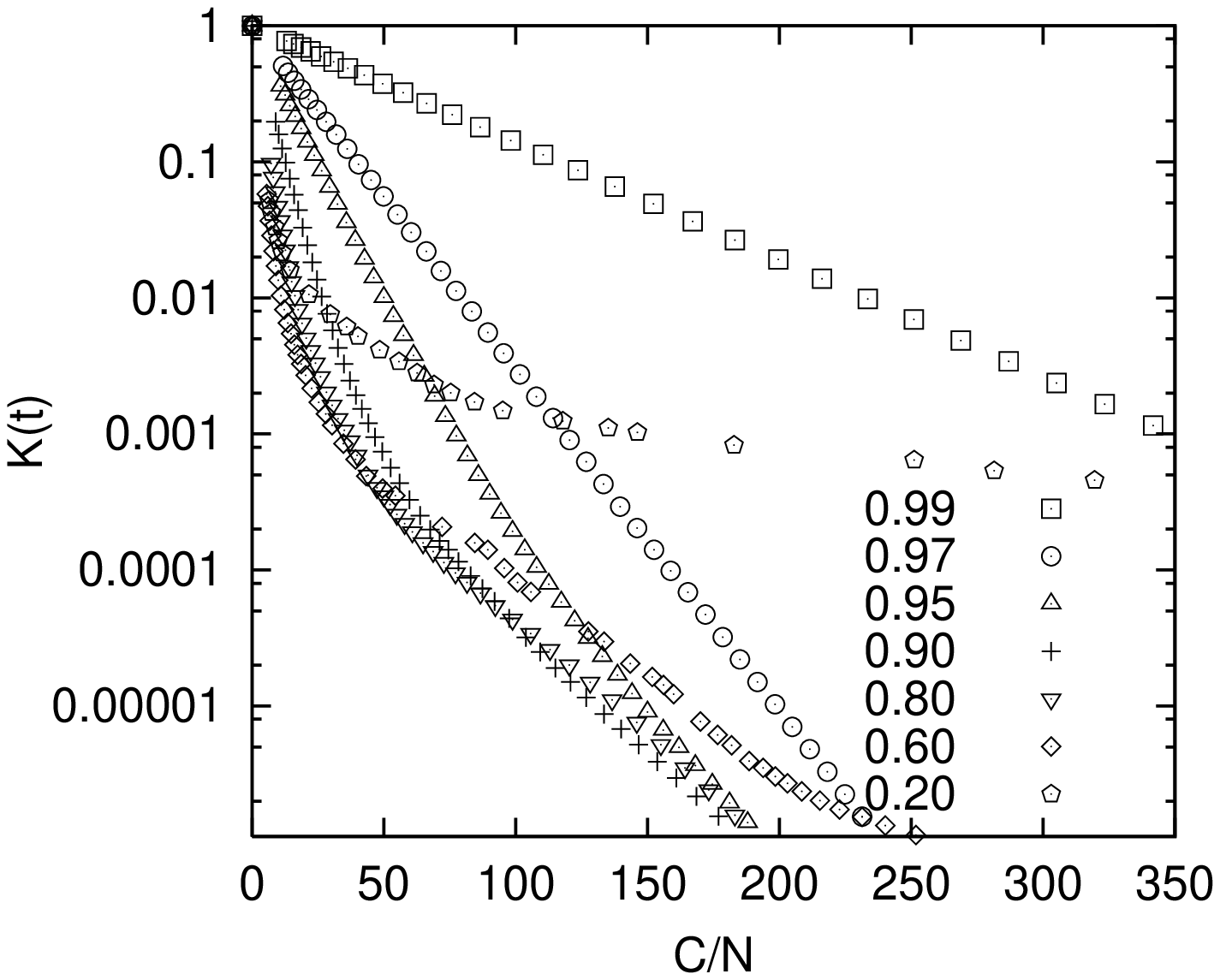,height=7.7cm,angle=-90} \hfill
 ~\hspace{-1cm}~ \epsfig{file=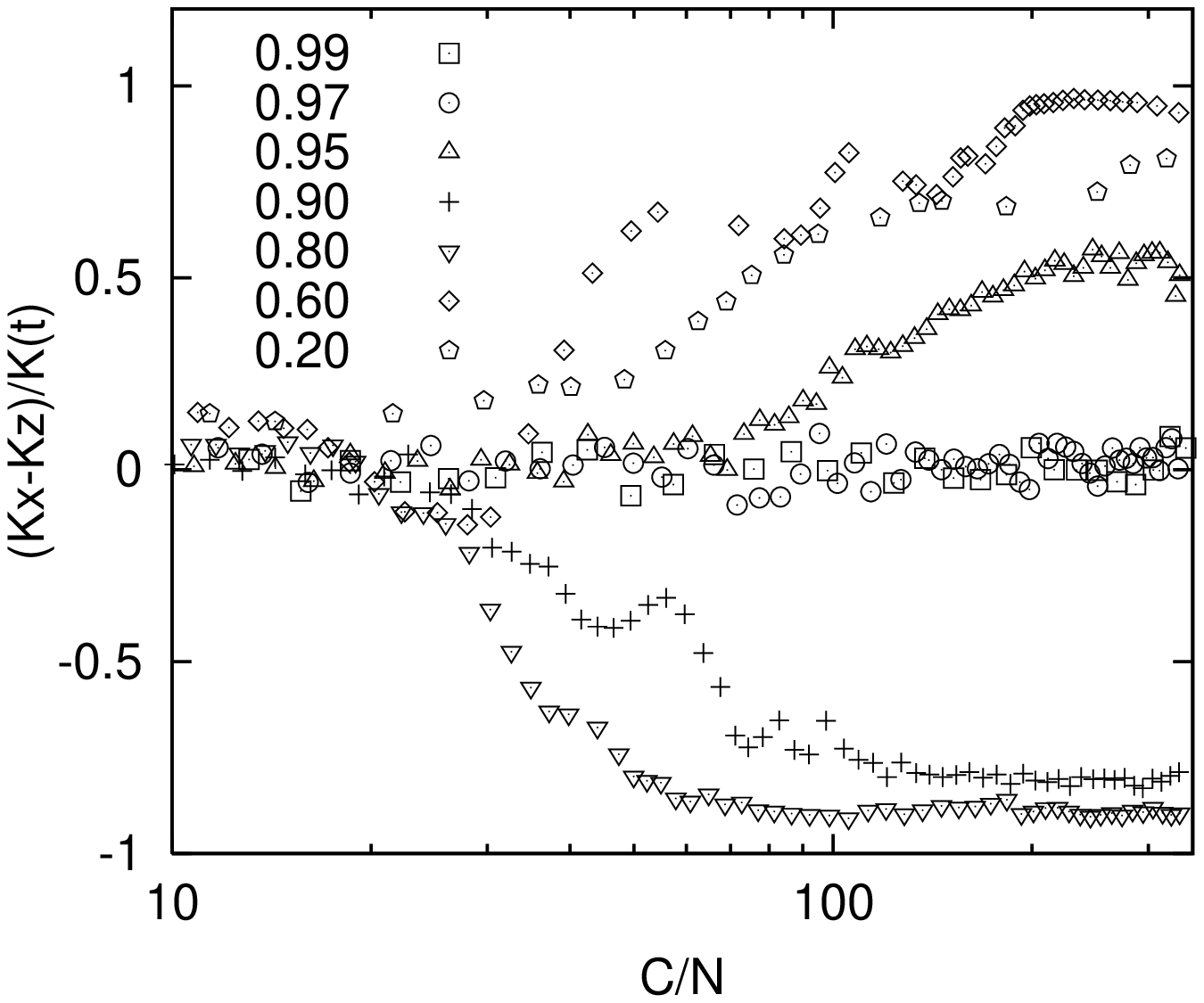,height=7.7cm,angle=-90} 
 ~\vspace{-.4cm} \\
\caption{
(a) $K(t)$ as function of $C/N$ with $t_c = 10^{-6}$\,s
and $r$ as given in the inset.
(b) $(K_x-K_z)/K(t)$ for the same simulations as in (a).
}
\label{fig:ene01}
\end{figure}

The restitution coefficient and the cut-off time $t_c$ are varied ($0.99 \le r \le 0.20$ 
and $10^{-10}\,{\rm s} \le t_c \le 10^{-3}\,$s) and the simulation is performed for 
each parameter set until every particle carried out $C/N = 1000$ collisions.
In Fig.\ \ref{fig:ene01}(a) $K(t)$ is plotted against the mean number of collisions 
per particle $C/N$ for simulations with $t_c = 10^{-6}$\,s and variable $r$. 
For large $r$ and small $C/N$ the energy behaves as  
$K(t) \propto \exp(-C/N)$ as can be derived from Eq.\ (\ref{eq:kt})
by integration over the product of mean velocity $\overline v = \sqrt{K/m}$ 
and inverse mean free path $\lambda \propto 1/\varrho$. 
With decreasing $r$ the initial slope is larger because more energy is dissipated
per collision. At larger times, energy decays much slower and $K(t)$ deviates
from the straight line already for $r \le 0.95$. In Fig.\ \ref{fig:ene01}(b)
$(K_x-K_z)/K(t)$ is plotted against $C/N$. In a homogeneous system without
clustering, the value of $(K_x-K_z)/K(t)$ fluctuates around zero, while
values close to unity indicate the `shearing-mode', i.e.~most of the kinetic
energy can be found in one direction \cite{mcnamara96b,goldhirsch96}. 
The deviations from the homogeneous cooling state begin earlier with decreasing $r$.

\begin{figure}[htb]
\vspace{.8cm}~
~~~~~~~~~~~~~~~~~~~~~~~(a) \hfill
(b)~~~~ \vspace{-1.6cm}\\
~\hspace{-1cm}~ \epsfig{file=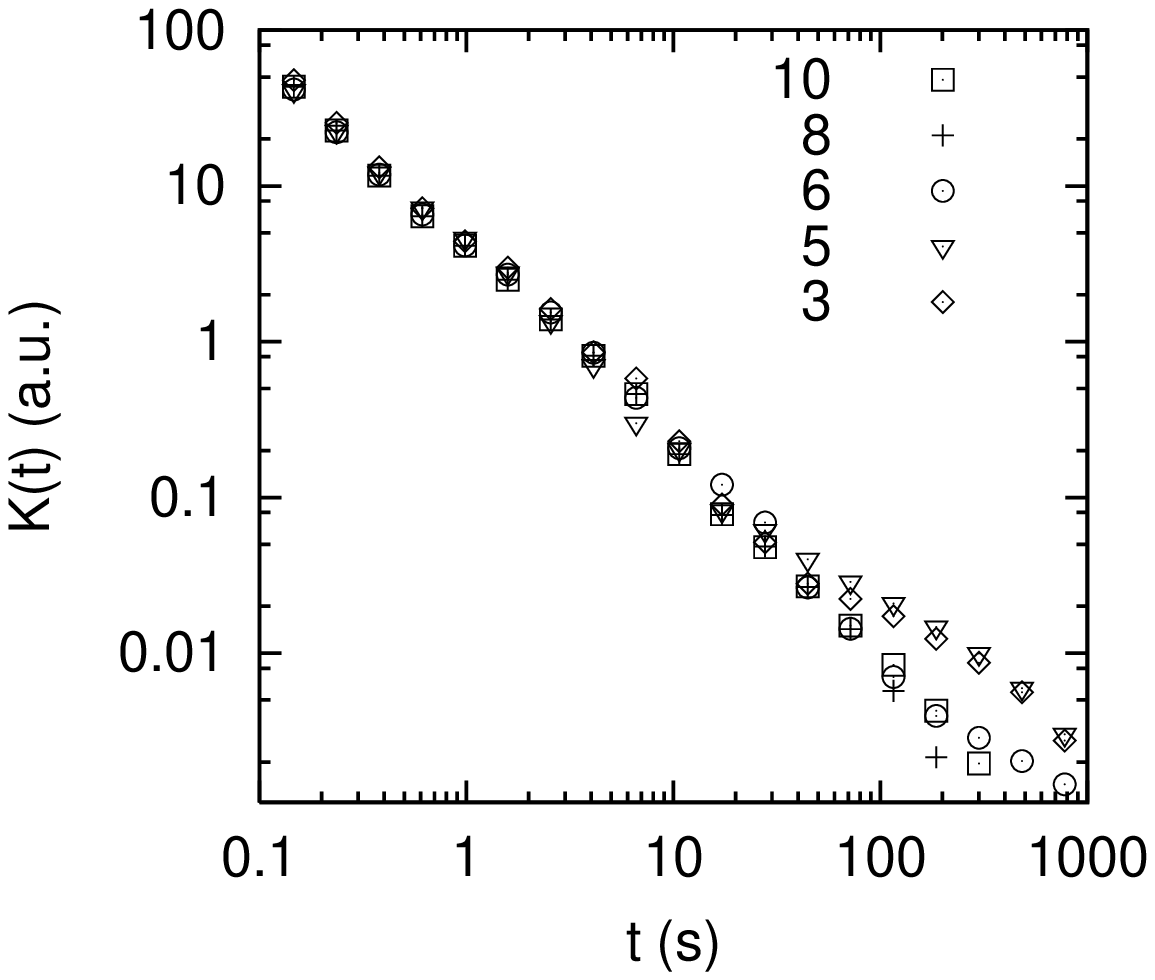,height=7.5cm,angle=-90} \hfill
~\hspace{-1cm}~ \epsfig{file=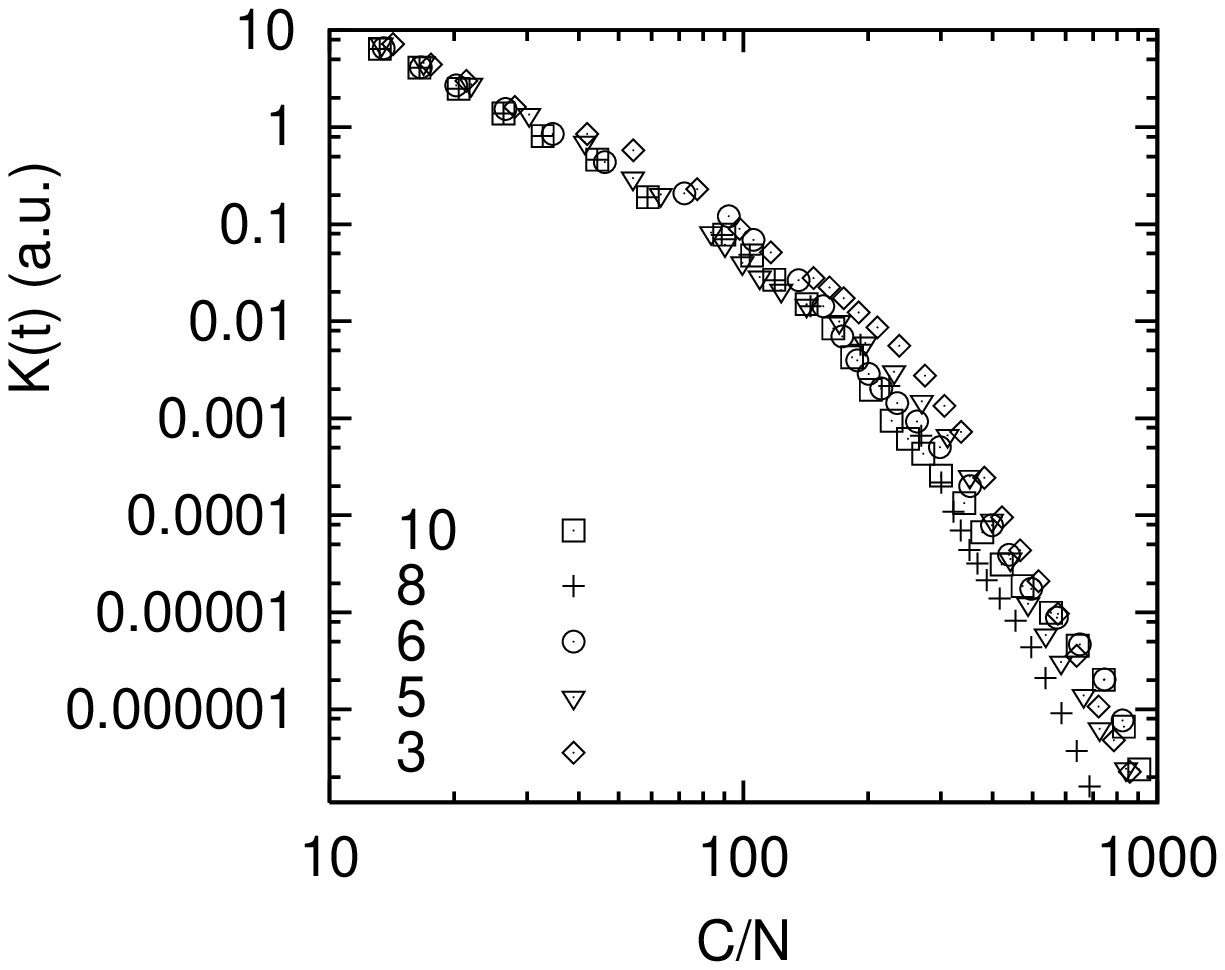,height=7.5cm,angle=-90} \\
\caption{
(a) $K(t)$ as function of $t$ with $r = 0.6$ and
$- \log_{10} t_c$ as given in the inset.
(b) $K(t)$ plotted against $C/N$ for the same simulations
as in (a).
}
\label{fig:ene02}
\end{figure}

For strong dissipation one observes jumps in $C/N$ --- meaning that
a few particles perform many collisions without affecting the global
energy too much. In that case one expects that the time between 
collisions drops below the threshold $t_c$ and $r$ is set to
unity according to Eq.\ (\ref{epsnew}). The kinetic energy $K(t)$ is 
plotted against $C/N$ in Fig.\ \ref{fig:ene02}(a) for 
simulations with $r =0.60$ and different $t_c = 10^{-10}$\,s,
$10^{-8}$\,s, $10^{-6}$\,s, $10^{-5}$\,s, and $10^{-3}$\,s.
For $t < 50$\,s the decay of energy is almost independent of
$t_c$. For larger times the simulations with greater $t_c$ loose
less energy since more collisions are elastic. From the plot
of $K(t)$ against $C/N$ in Fig.\ \ref{fig:ene02}(b) one observes
for $C/N < 100$ a power-law behavior $K(t) \propto (C/N)^\zeta$ 
with $\zeta \approx 2.5$ and for $C/N > 100$ a much faster decay 
of energy, with $\zeta \approx 7$.

\subsection{Cluster growth}

\begin{figure}[ht]
\vspace{.8cm}~
~~~~~~~~~~~~~~~~~~~~~~~~~~~~~~~~~~~~~~~~~~(a) \hfill
(b)~~~~~~~~~~~~~~~~~~~~~~~~~~~~~~~~~~~~~~~ \vspace{-1.6cm}\\
~\hspace{-1.cm}~ \epsfig{file=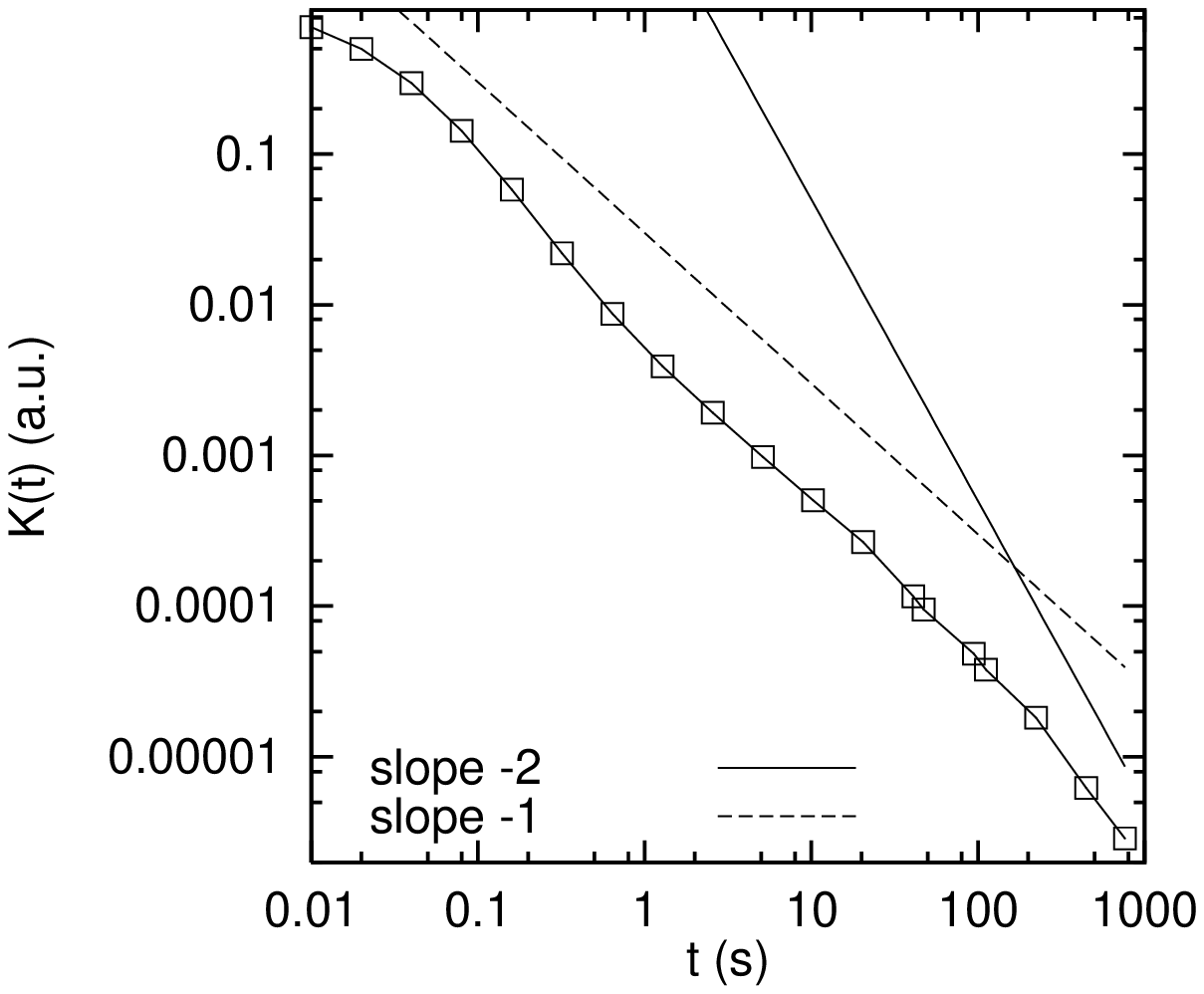,height=7.5cm,angle=-90} \hfill
~\hspace{-1.cm}~ \epsfig{file=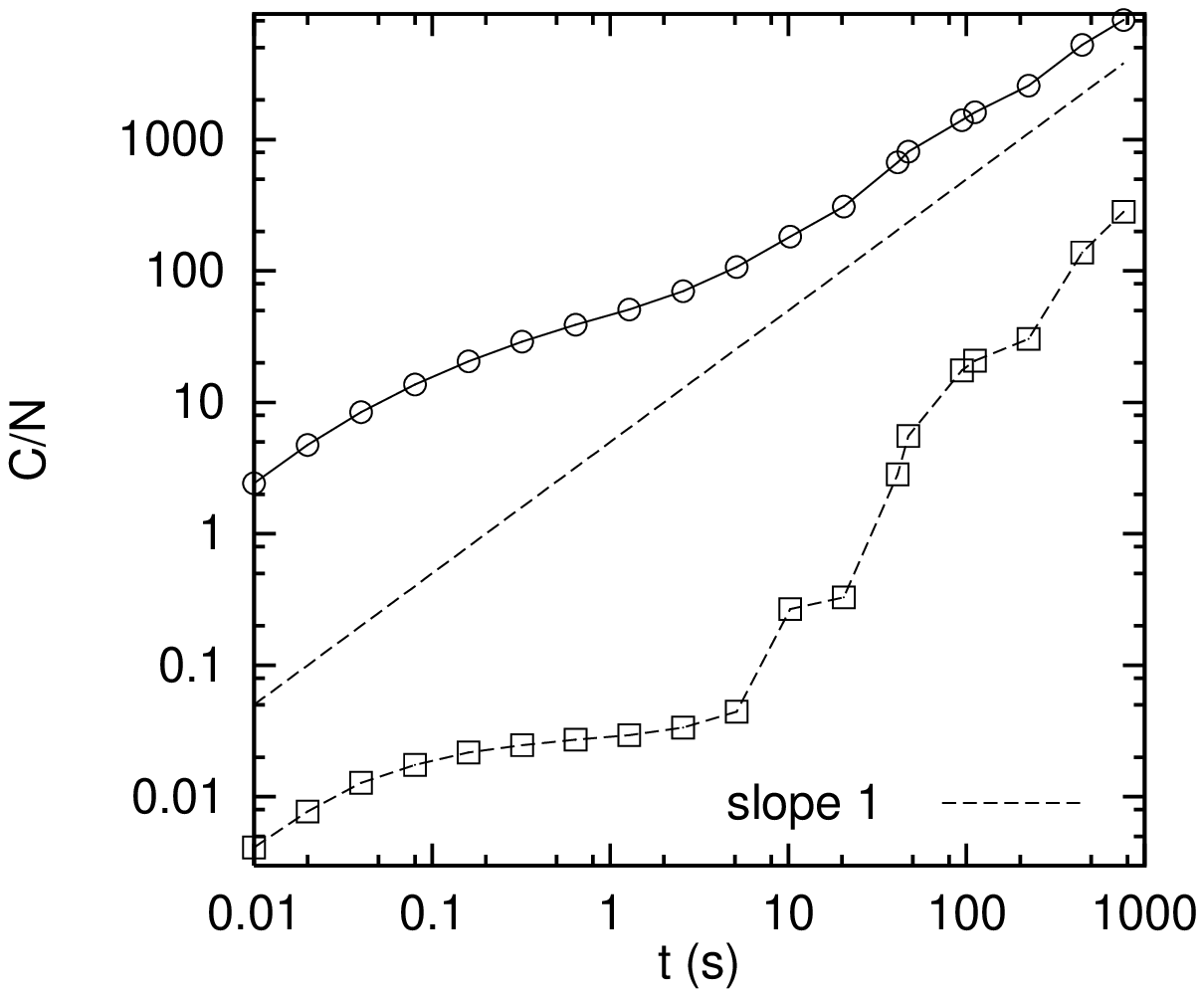,height=7.5cm,angle=-90} 
\caption{
(a) $K(t)$ as function of time $t$ for a simulation
with $N=79524$, $L=500$, $\varrho = 0.25$, and $t_c = 10^{-5}$\,s.
(b) $C/N$ as function of $t$ for the simulation in (a). The circles
give the mean number of collisions per particle and the squares give 
the mean number of elastic collisions per particle.
}
\label{fig:cluster_K}
\end{figure}

In the  following we discuss a simulation with 
$N=79524$ particles, $L=500$, $\varrho = 0.25$, and
$t_c = 10^{-5}$\,s in more detail. In Fig.\ \ref{fig:cluster_K}(a)
the energy $K(t)$ is plotted against the simulation time.
The solid line indicates a slope of $-2$ as one would get in the 
homogeneous cooling regime, see Eq.\ (\ref{eq:kt}). However, the
simulation seems to follow, in average, a slope of -1, as indicated 
by the dashed line. Thus the cooling is much slower when 
it is non-homogeneous. Fig.\ \ref{fig:cluster_K}(b)
shows that the fraction of particles that have collisions with
a collision frequency larger than $1/t_c$ is only about
0.1 percent of the total number of particles for $t < 50$\,s.
For larger times, the number of elastic collisions increases, but
never above three percent of $C/N$.

\begin{figure*}[htbp]
{$t=0.640$ s, $C/N=39$} \hfill {$t=10.24$ s, $C/N=183$} \\
\epsfig{file=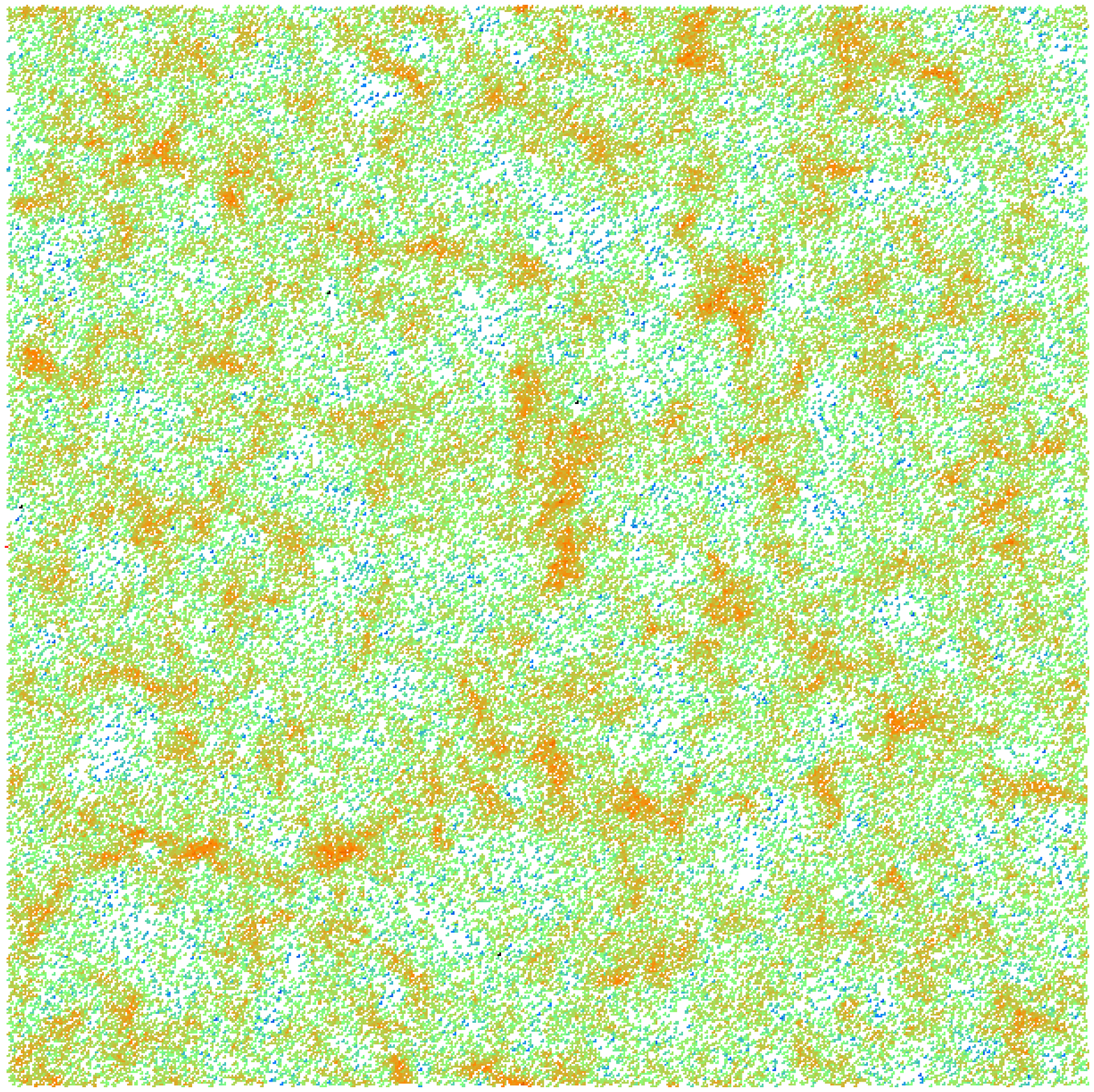,height=7.7cm} \hfill
\epsfig{file=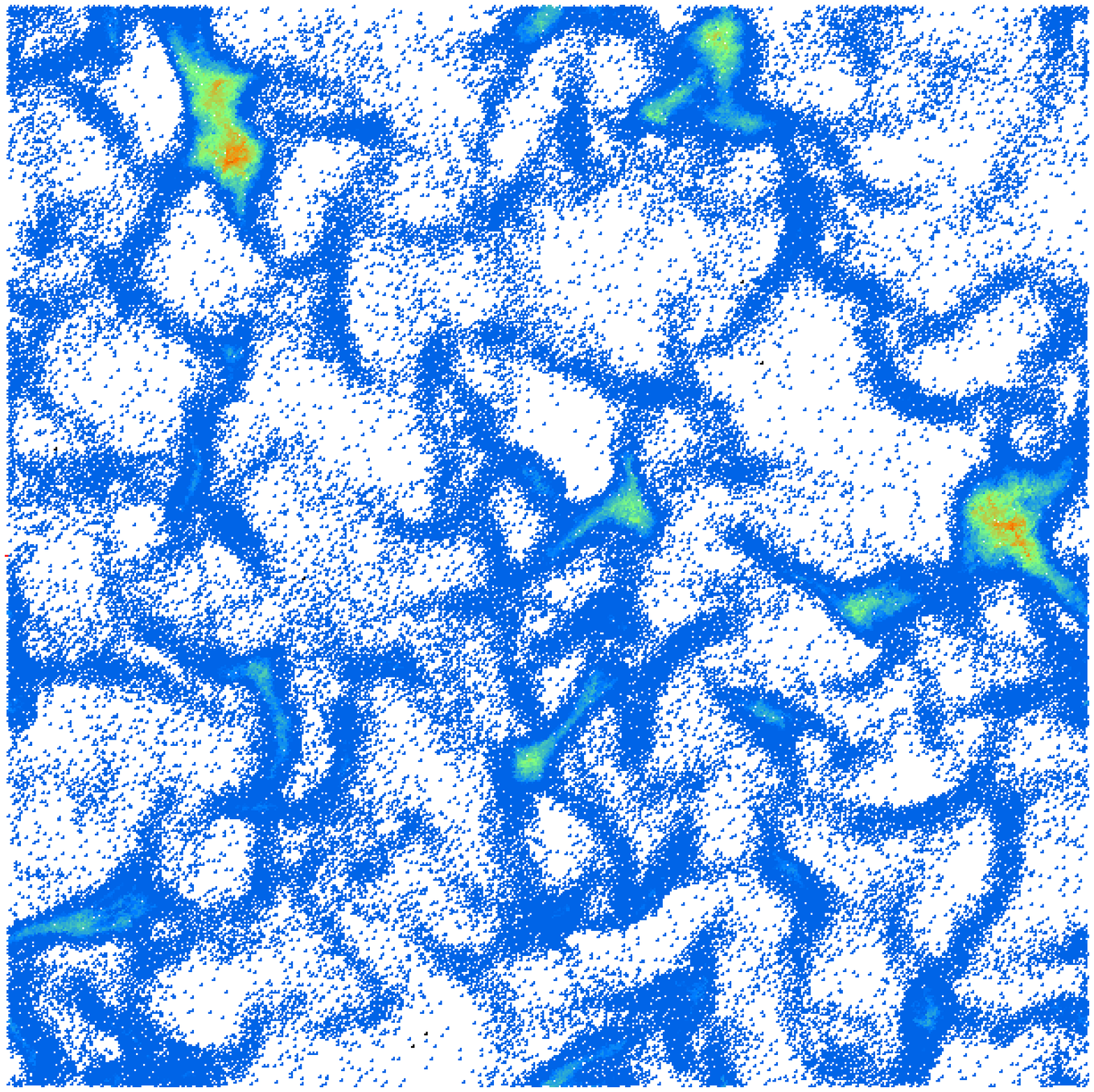,height=7.7cm}\\
{$t=223.2$ s, $C/N=2567$} \hfill {$t=446.6$ s, $C/N=5258$} \\
\epsfig{file=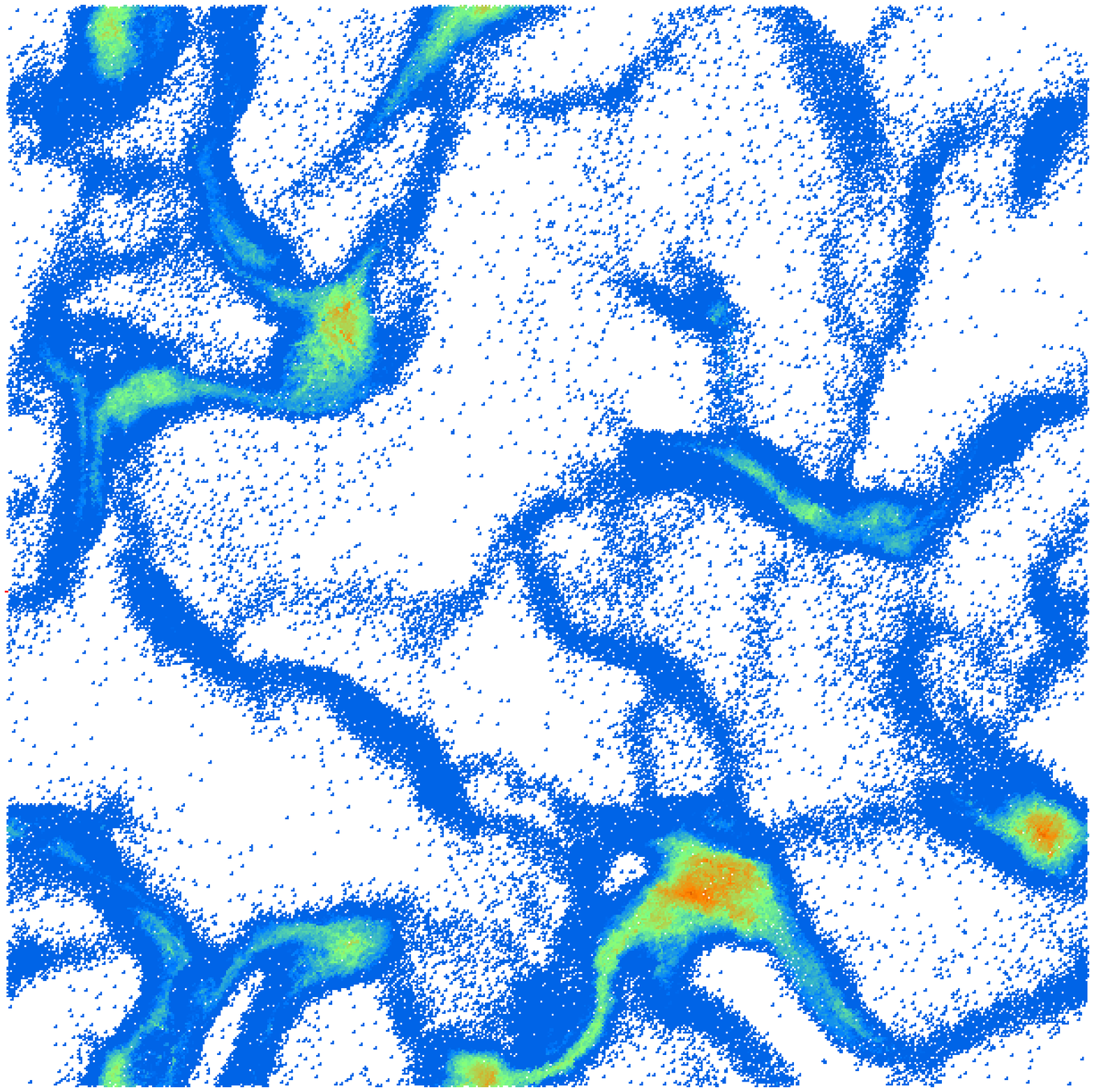,height=7.7cm} \hfill
\epsfig{file=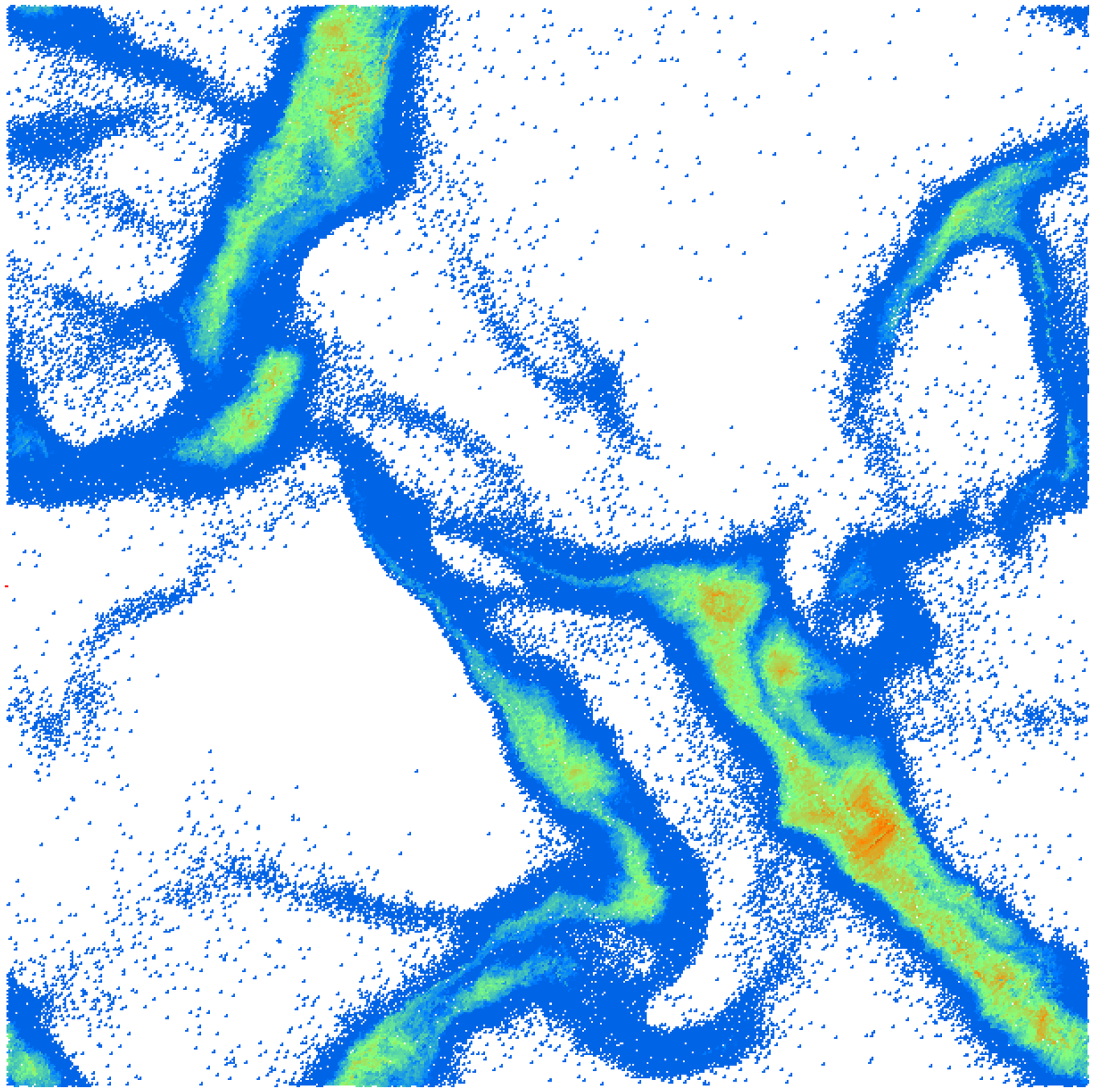,height=7.7cm}\\
\caption{ED simulation with $N=79524$ particles in a system of 
size $L=500$, volume fraction $\varrho = 0.25$, restitution coefficient 
$r=0.8$, and critical collision frequency $1/t_c = 10^5$ s$^{-1}$.
The collision frequency is color-coded --- red and blue correspond 
to small and large collision frequencies, respectively.
}
\label{fig:cluster_nc}
\end{figure*}

In Fig.\ \ref{fig:cluster_nc} snapshots of the simulation in Fig.\ 
\ref{fig:cluster_K} are displayed. With increasing time $t$ (and $C/N$) 
structures build up in the system and grow in size. In the bright regions 
in the centers of the clusters the collision frequency is largest.

\subsection{Probability distribution of the collision frequency}

\begin{figure*}[ht]
\begin{tabular}{ p{7.4cm} p{0.4cm} p{7.4cm} }
~\hspace{-0.2cm} \epsfig{file=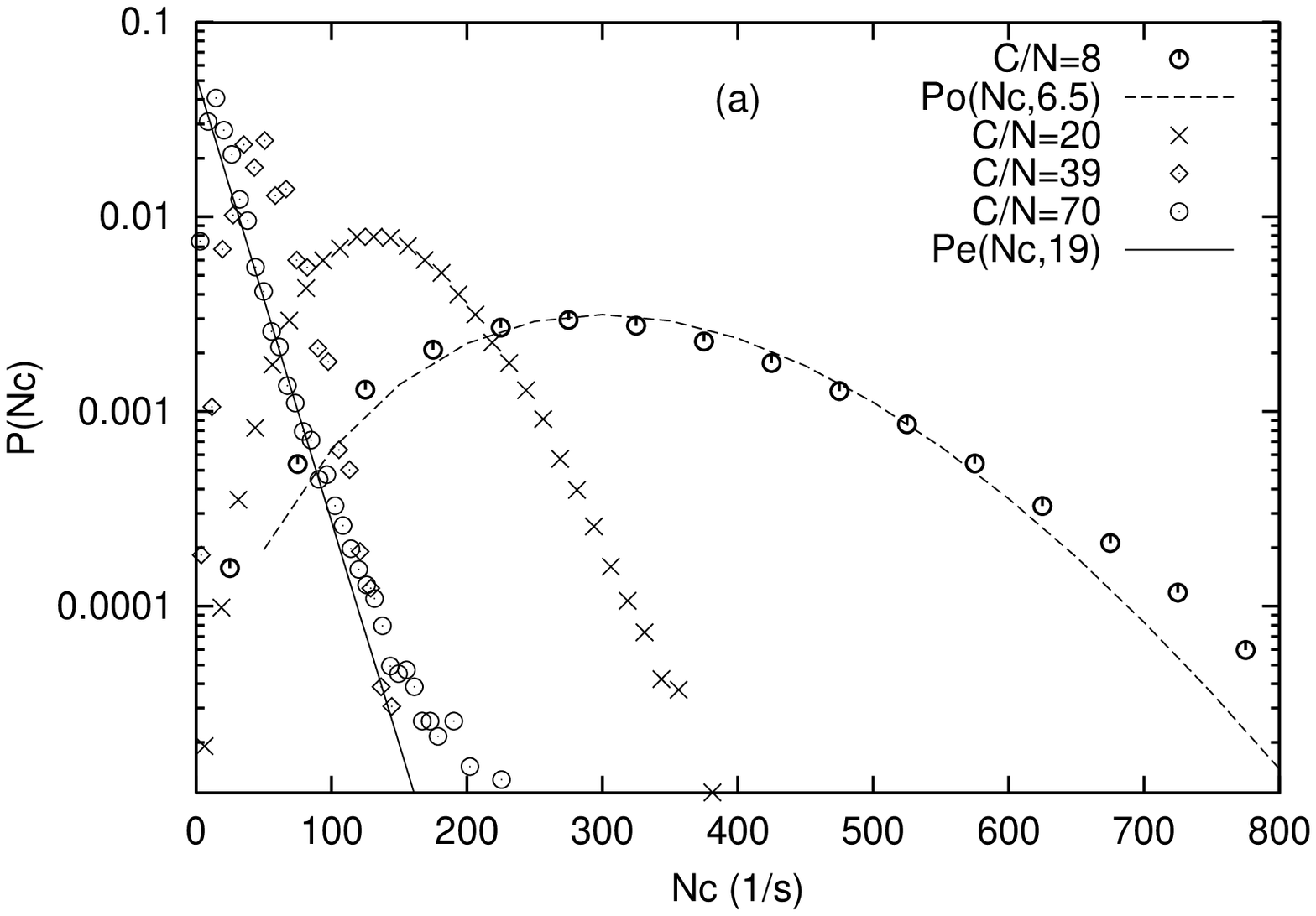,height=7.6cm,angle=-90} & &
~\hspace{-0.4cm} \epsfig{file=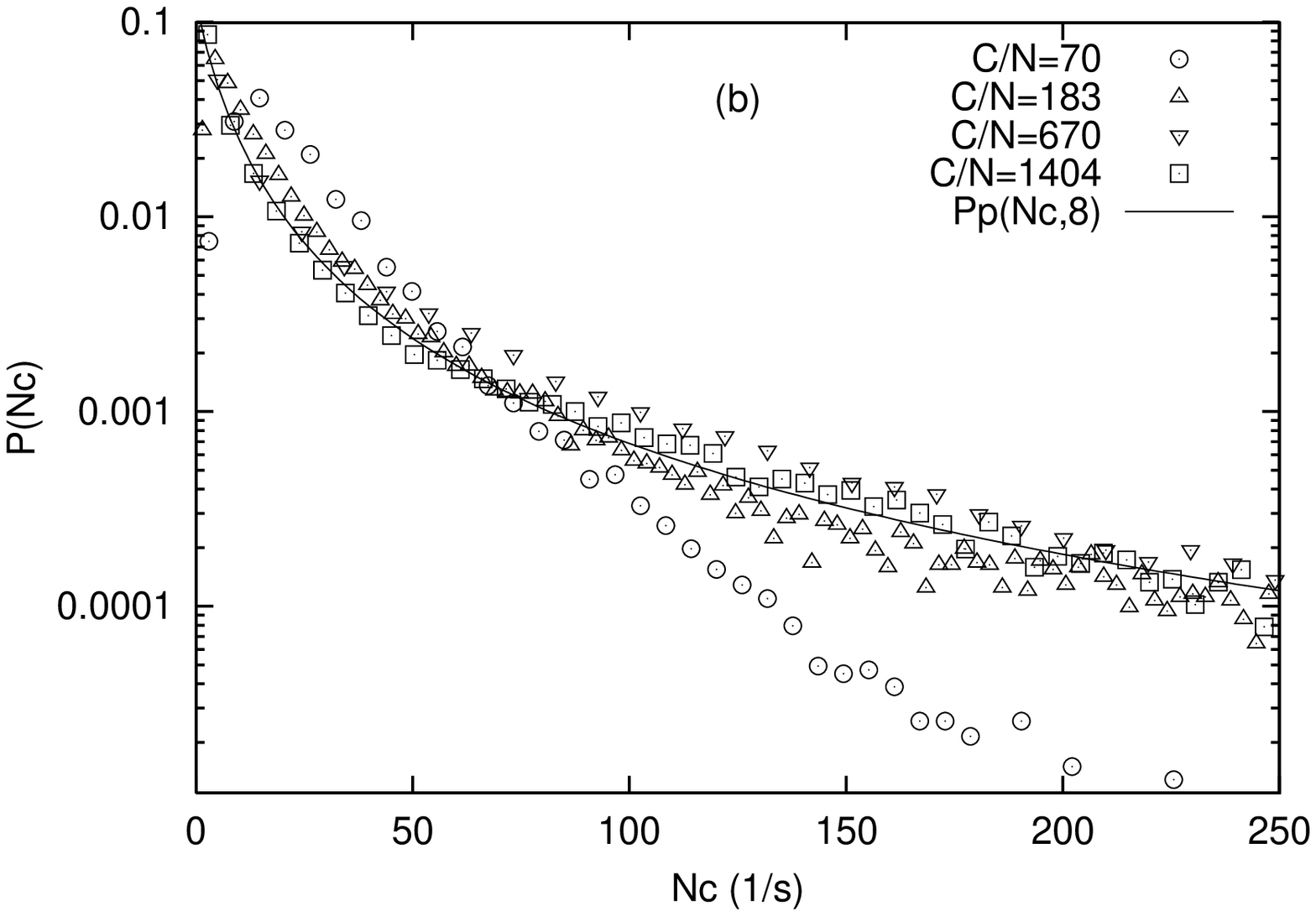,height=7.6cm,angle=-90} \\
~\hspace{-0.2cm} \epsfig{file=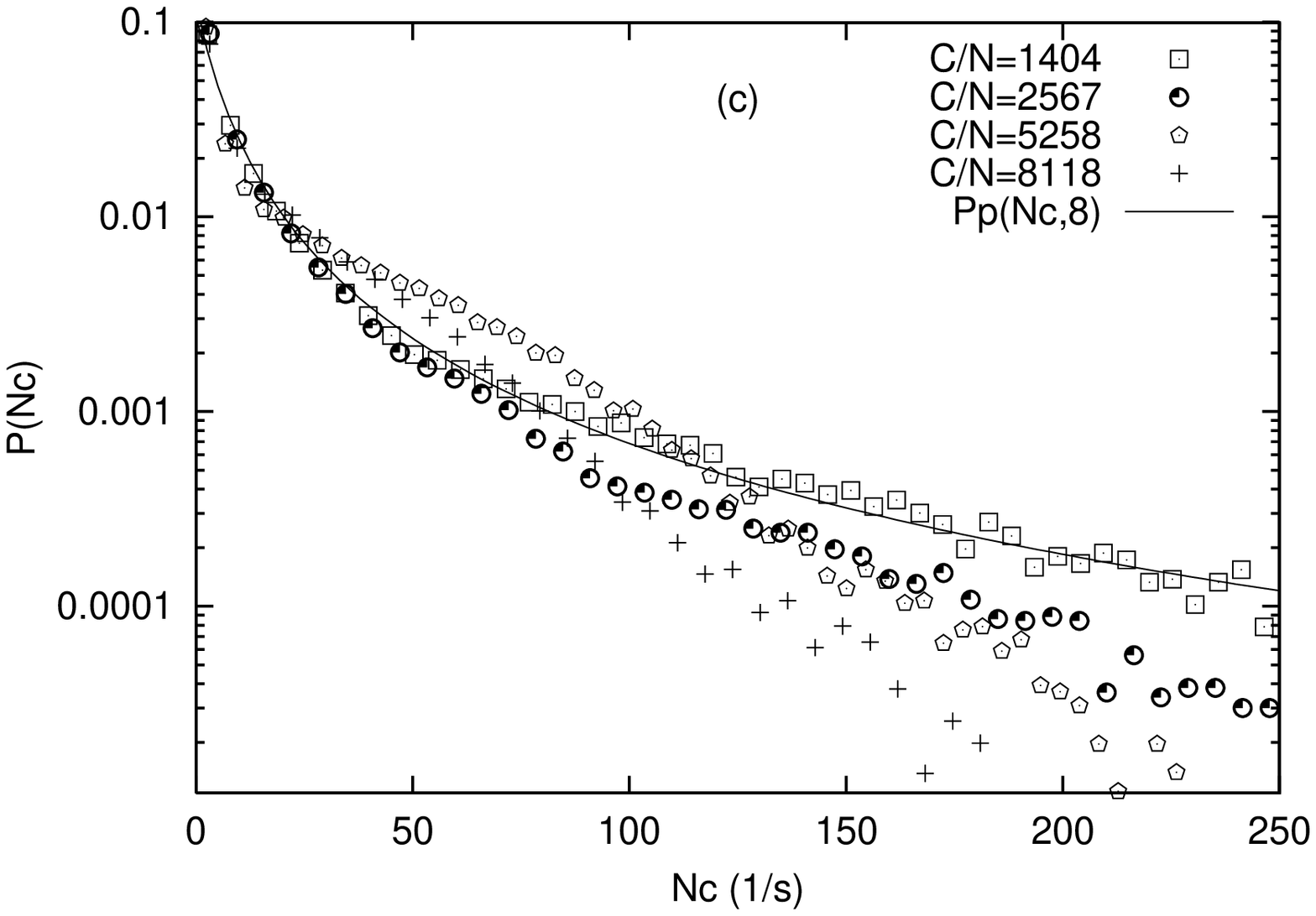,height=7.6cm,angle=-90} & &
~\vspace{0.7cm}~
\caption{Semi-log plot of
$P(N_c)$ for different $C/N$ values from the simulation displayed in Figs.\
\protect\ref{fig:cluster_K} and \protect\ref{fig:cluster_nc}. The system is
(a) in the homogeneous cooling state, (b) in the cluster growth regime, and
(c) in the long-time state when the largest cluster has reached the system
size.
}
\label{fig:nch}
\end{tabular}
\end{figure*}

For a more quantitative analysis of the clustering,
the probability distribution for particle collision 
frequencies is examined. $P(N_c)$
gives the probability to find a particle
that carried out $N_c$ collisions per unit time in the last 
time interval $\Delta t = t/2$.
In Fig.\ \ref{fig:nch}(a) $P(N_c)$ is plotted for the homogeneous 
cooling during short times. The shape of $P(N_c)$ resembles a Poisson 
distribution whose mean and width $\chi/\Delta t$ continuously
decrease with time for $C/N \le 39$. For $C/N=8$, i.e.~$t=0.04$\,s,
the dashed line corresponds to $P_o({\cal C}/\Delta t,6.5)\,\Delta t$ with the
Poisson distribution for the number of collisions ${\cal C}$ per particle per
time $\Delta t$
\begin{equation}
P_o({\cal C},\chi) = { \exp(-\chi) \chi^{\cal C} \over {\cal C} ! } ~,
\end{equation}
with mean and width $\chi/\Delta t = 6.5/0.02\,$s$ = 325$\,s$^{-1}$.
This distribution indicates that the collisions are uncorrelated events at
the beginning of the simulation.

At longer times, the probability for a large number of collisions 
can be well fitted by an exponential 
\begin{equation}
P_e(N_c,A)= \frac{1}{A} \exp \left (- \frac{N_c}{A} \right ) ~,
\label{eq:pfig1}
\end{equation}
with the mean collision rate $A=19$~s$^{-1}$ in Fig.\ \ref{fig:nch}(a).
As soon as clustering occurs, the form of the probability of the number
of collisions changes dramatically. 
In Figs.\ \ref{fig:nch}(b) and (c), $P(N_c)$ is
displayed at different times, i.e.~different $C/N$.
The probability for large (small) collision frequencies
increases (decreases), and can now be approximated by
a power-law of the form
\begin{equation}
P_p(N_c,B)= \frac{B}{\left ( B+{N_c} \right ) ^2} ~,
\label{eq:pfig2}
\end{equation}
with the rate $B=8$~s$^{-1}$.
All functions $P(N_c)$ are normalized for arbitrary $\chi$, $A$ or $B$, but 
$P_p(N_c,B)$ has a power-law tail so that the first moment diverges. 
This indicates the inelastic 
collapse, i.e.~the divergence of the global collision frequency. However,
in the simulation presented here, the collision frequency is limited 
since $t_c$ is non-zero. 

In summary, we found that for $C/N > 70$ the shape of $P(N_c)$ changes from an
exponential decay for large $N_c$ to a power-law behavior. This corresponds to 
the density instability when clusters evolve and grow with time, 
i.e.~cooperative motion and interaction.
Interestingly, the shape of the probability distribution is
only weakly varying over one order of magnitude in time, from $C/N \approx 183$ 
to $C/N \approx 1404$. At these times, particles that carry out
many collisions coexist with those which carry out only a few.
For even larger times $C/N > 2567$ the function $P(N_c)$ changes
shape again, as displayed in Fig.\ \ref{fig:nch}(c); the probability
for large collision frequencies drops again. 

\section{Simulating the Flow through Pipes}
\label{sec:pipeflow}

In this section we present another collective phenomenon
in a totally different geometry. 
Recent observations of approximately V-shaped {\it microcracks }
in vertically vibrated sand-piles \cite{duran94} led 
to the problem of gravity driven vertical motion of sandpiles in 2D
pipes \cite{duran96,luding96,luding96b,luding96c}. In this situation
the material is accelerated by the gravitational force and confined
to the pipe by two side walls. During the fall, cracks develop
in the lower part of the pile and ascend progressively inside the
bulk in both experiment and simulation \cite{duran96}.
For details on experiments and simulations see Refs.\ 
\cite{duran96,luding96b}.

The reasons for a crack to open have been 
identified as the fluctuations of either the wall surface \cite{duran96}
or the random particle motion \cite{luding96b}. It was reported that 
fluctuations lead to a momentum wave in the material.
A part of the material is decelerated and the material from above
hits this slower plug and causes a new, possibly stronger momentum wave.
The increased pressure on the sidewalls may lead to an even stronger 
deceleration so that eventually a crack opens below the plug
--- and becomes visible only that late.
Thus, cracks are a rather bad indicator of the dynamical processes
occuring inside a granular material, since they need some time
to open, before they get visible. In simulations one has access
to quantities that allow deeper insight what is going on in 
the material. The number of collisions per unit time in which a particle
participates can be measured and visualized easily. 

\begin{figure*}[htb]
~\hspace{-0.8cm}~
~\vspace{-0.4cm}~
\epsfig{file=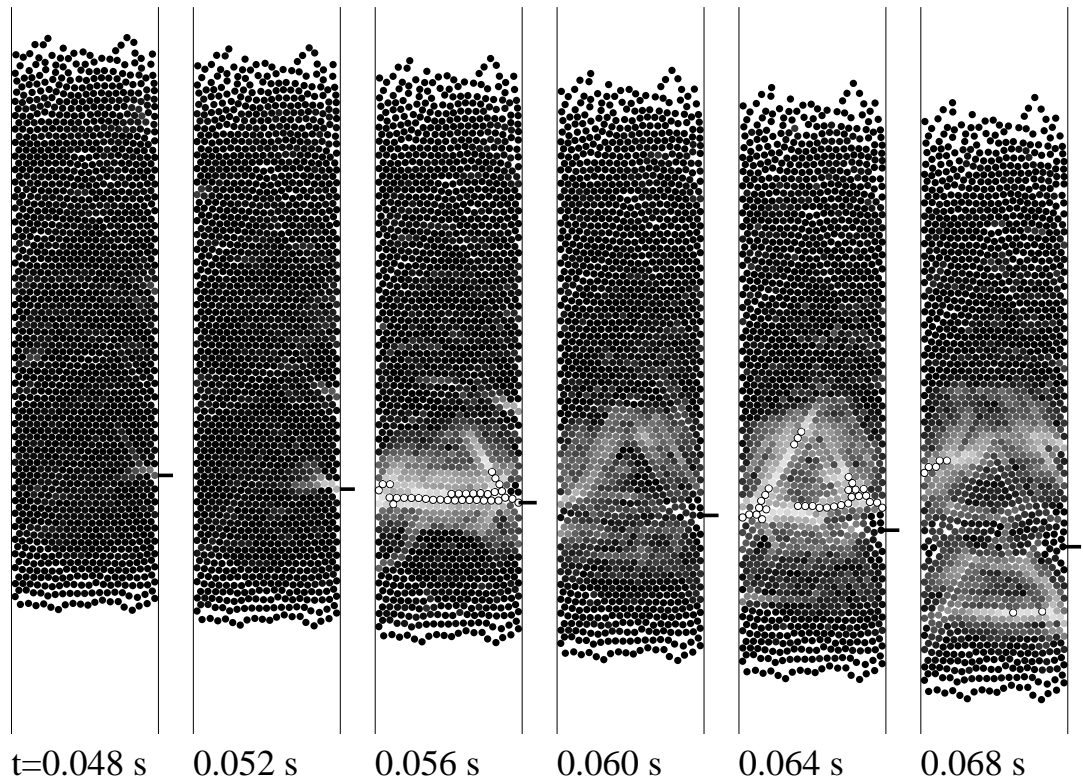,height=11.4cm}
\caption{Snapshots from a simulation with
$N = 1562$, $L = 20.2$, $\epsilon = \epsilon_w = 0.9$,
$\mu = \mu_w = 0.5$, and $\beta_0 = \beta_{0w} = 0.2$.
Black particles have a collision frequency of $N_c = 0$\,s$^{-1}$, white
particles have $N_c \ge 10^4$\,s$^{-1}$, and
grey particles interpolate between the two extremes.}
\label{fig:shock}
\end{figure*}

The system is a box of width $L=l/d$ and initially filled with $N$
particles with diameter $d$, situated on a triangular lattice
with lattice constant $s=1.01\,d$. Each particle is assigned a random velocity,
uniformly distributed in the range $-v_0\le v_i(0)\le v_0$ in both, the
horizontal and vertical directions. This rather regular system is now allowed
to reach a steady state, i.e.~we start the simulation at $t=-t_r$, using $%
r =r _w=1$ and $\mu =\mu _w=0$. A typical average velocity in
our simulations is 
$\overline{v}=\sqrt{<v^2>}$ = 0.05 \,m\,s$^{-1}$ for $t = 0$\,s.
Due to the rather low kinetic energy, the array of particles is still arranged
on a triangular lattice, except for a few layers at the top which are
fluidized. In a typical simulation, we use $L = 20.2$ and $N = 1562$, 
so that the array consists of about 80 layers.
At time $t=0$\,s we remove the bottom, switch on dissipation and friction and
let the array fall. In Fig.\ \ref{fig:shock} we present data from one special 
simulation with $N = 1562$, $L = 20.2$, $\epsilon = \epsilon_w = 0.9$, 
$\mu = \mu_w = 0.5$ and $\beta_0 = \beta_{0w} = 0.2$.
Light (dark) particles carried out
many (few) collisions in the last interval $\Delta t = 0.001$~s.

The black bar at the right wall marks the particle at which
one observes at first a large collision frequency $N_c$
at time $t = 0.048$~s.
Already $0.004$~s later the neighboring particles react and
also carry out more and more collisions. This increase in 
collision frequency leads to an increase of pressure that, 
in return, leads to more friction, a slow-down of the particles
close to the wall, and to even more collisions with
the following particles. The grey region indicates an arch-like 
structure that spans the whole width of the system
at $t = 0.056$~s. Again $0.004$~s later this region of
large $N_c$ and large pressure has almost disappeared, and
for $t \ge 0.060$~s at the position of the initiating particle
(at the black bar) a crack is visible. Even later, new arches 
above and below the original one appeared.

\begin{figure*}[htb]
~\hspace{-0.2cm}~ \epsfig{file=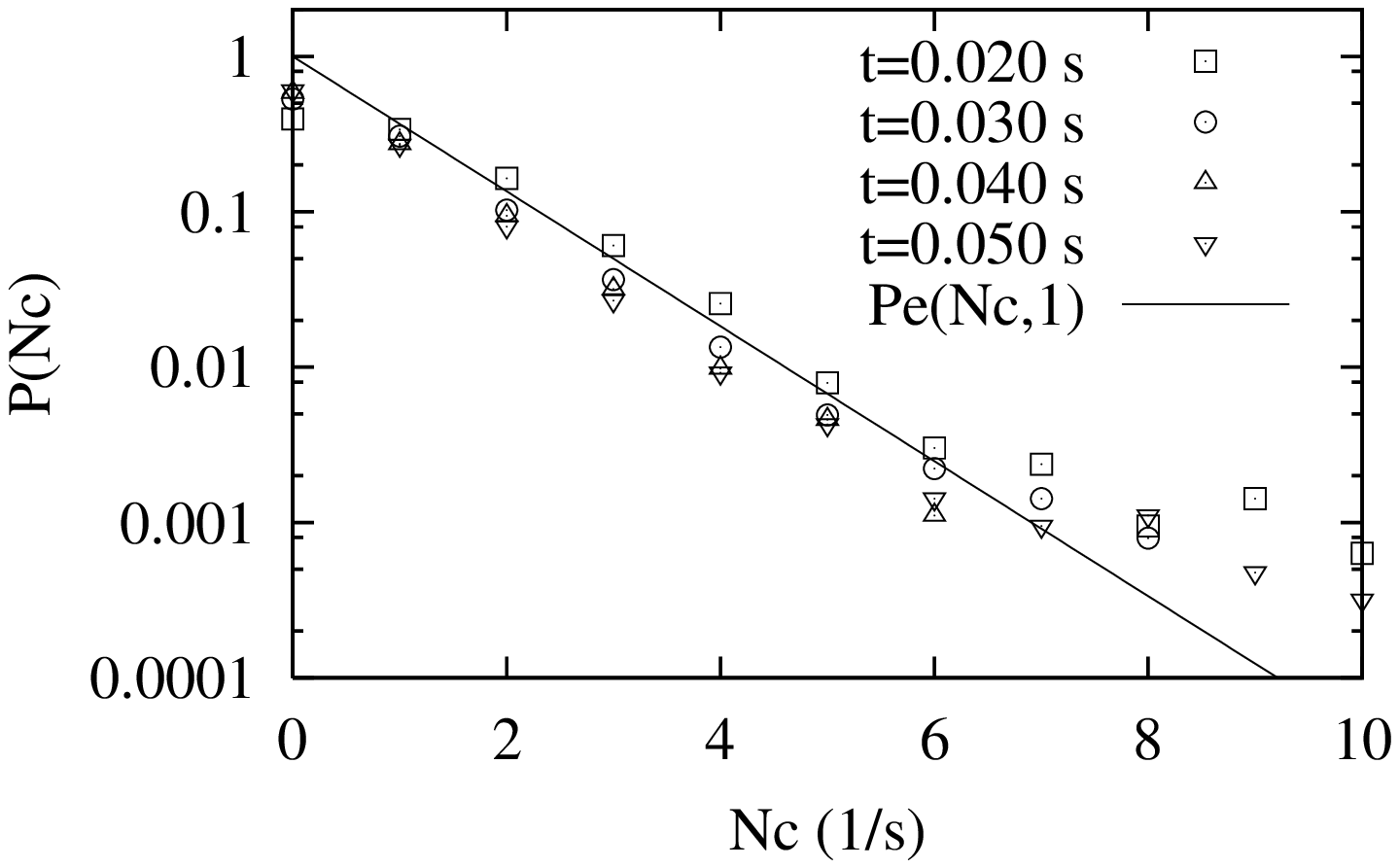,height=4.6cm}
~\vspace{ 0.5cm}~ \epsfig{file=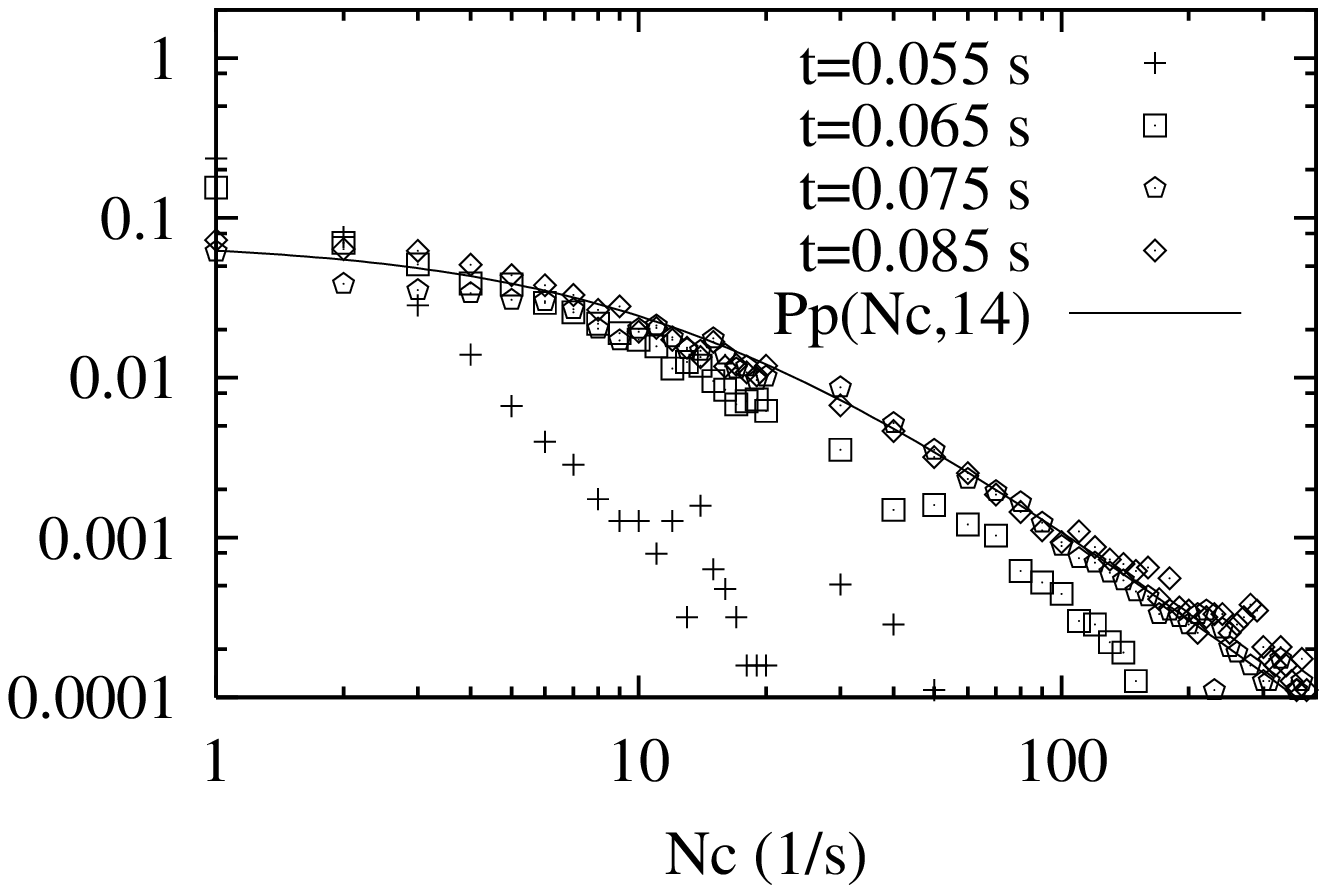,height=4.6cm}
\caption{
(a) Semi-log plot of the probability distribution
$P(N_c)$ against the number of collisions per unit time
from the simulation in Fig.\ \ref{fig:shock}. The data
correspond to a time before cracks and arching occurs,
the line gives $P_e$ from Eq.\ (\protect\ref{eq:pfig1}) with
$A=1$.
(b) Log-log plot of $P(N_c)$ from the same simulation as
in (a) at later times, when cracks and arches exist.
The solid lines gives Eq.\ (\protect\ref{eq:pfig2})
with $B=14$.
}
\label{fig:pfig}
\end{figure*}

In Fig.\ \ref{fig:pfig} we present again the probability $P(N_c)$
to find a particle in the system that performed a number of
$N_c$ collisions per unit time within the last time interval 
$\Delta t = 0.005$~s.
$P(N_c)$ is calculated from the simulation in Fig.\ \ref{fig:shock}
at times $t = 0.02$~s, 0.03~s, 0.04~s, and 0.05~s, i.e.~before the
first arch (or crack) occurs. The probability for the number 
of collisions can be well fitted by the exponential $P_e(N_c,A)$,
see Eq.\ (\ref{eq:pfig1}),
with the mean collision rate $A=1$~s$^{-1}$.
As soon as arches occur, the probability of the number
of collisions changes shape. In Fig.\ \ref{fig:pfig}(b), $P(N_c)$ is
displayed at times $t = 0.055$~s, 0.065~s, 0.075~s, and 0.085~s.
The probability for large (small) collision frequencies
increases (decreases), and can now be approximated by
the power-law $P_p(N_c,B)$
with the rate $B=14$~s$^{-1}$ from Eq.\ (\ref{eq:pfig2}).
Note that the curves in Fig.\ \ref{fig:nch} and Fig.\ \ref{fig:pfig}
are both Eq.\ (\ref{eq:pfig2}), however, with different $B$.
They look different because the horizontal axis is
linear in the first but logarithmic in the second figure.

\section{Summary and Conclusion}
\label{sec:summary}

With a rather simple description of a granular material as an 
ensemble of inelastic spherical particles we have investigated
interesting effects like the clustering instability and arching. 
A freely cooling granulate builds up
a density instability and clusters of particles are formed.
The structure on small length scales, catched by 
the event driven simulation, could not be reproduced by
the stochastic DSMC method. The large scale structures, i.e.~the
structure factor of the clusters, however, was nicely reproduced
with the stochastic method. 

DSMC assumes molecular chaos. Thus we examined different systems 
and found in 2D a constant probability distribution for the 
impact parameter, i.e.~molecular chaos, when the system is elastic
or slightly inelastic. Only for large densities, large systems and
strong dissipation, the molecular chaos assumption becomes invalid.
In those cases, we observed a shearing motion of grains, a phenomenon
that can be found also in large, dilute systems after dense clusters
have formed. DSMC cannot model dense clusters and also ED has
problems to handle dense regions with large collision frequencies.
Therefore, we introduced the advanced TCED method that avoids the
inelastic collapse, i.e.~dissipation is switched off when too 
much collisions occur per unit time. We found that this variation
of the traditional ED method involves only a small percentage of
the particles and thus does not affect the system's behavior 
as long as the cut-off time $t_c$ is small enough.

Finally, the statistics of the particle collision frequencies was
examined for two totally different boundary conditions.
In both cases, freely cooling and pipe flow 
as well, the probability distribution of the collision frequencies
shows two types of behavior. In the homogeneous, random regime
the distribution resembles a Poisson distribution (i.e.~an exponential 
for small mean values), indicating that the collisions are uncorrelated
events. 

As soon as cooperative effects like clustering or arching occur,
the probability distribution changes from an exponential to a 
power-law shape. We proposed {\em one} functional form that approximated 
the distribution functions for {\em both} boundary conditions equally well.
The described cooperative phenomena are of local nature but leave a 
fingerprint, i.e.~a power-law, in the global distribution function 
of the collision rate.
An open issue is the theoretical verification and understanding
of the shape of this interesting probability distribution function. 

\section*{Achnowledgements}

We thank E. Cl\'ement, and S. Weinketz for helpful discussions, and
are especially grateful to M. M\"uller for the DSMC data. Furthermore,
the financial support of the Deutsche Forschungsgemeinschaft (SFB 382)
is acknowledged.

%\bibliographystyle{unsrt}
%\bibliography{/home/lui/LIT/granulates}

\begin{thebibliography}{10}

\bibitem{wolf97}
D.~E. Wolf and P.~Grassberger, editors.
\newblock {\em Friction, Arching and Contact Dynamics}.
\newblock World Scientific, Singapore, 1997.

\bibitem{behringer97}
R.~P. Behringer and J.~T. Jenkins, editors.
\newblock {\em Powders \& Grains 97}.
\newblock Balkema, Rotterdam, 1997.

\bibitem{luding98}
S.~Luding.
\newblock {\em Die {P}hysik trockener granularer {M}edien}.
\newblock Logos Verlag, Berlin, 1998.
\newblock Habilitation thesis (in german),~ISBN 3-89722-064-4.

\bibitem{herrmann98}
H.~J. Herrmann, J.-P. Hovi, and S.~Luding, editors.
\newblock {\em Physics of dry granular media - NATO ASI Series E 350}.
\newblock Kluwer Academic Publishers, Dordrecht, 1998.

\bibitem{baxter89}
G.~W. Baxter, R.~P. Behringer, T.~Fagert, and G.~A. Johnson.
\newblock Pattern formation in flowing sand.
\newblock {\em Phys. Rev. Lett.}, 62(24):2825, 1989.

\bibitem{duran94b}
J.~Duran, T.~Mazozi, E.~Cl\'ement, and J.~Rajchenbach.
\newblock Decompaction modes of a two-dimensional ``sandpile'' under vibration:
  model and experiments.
\newblock {\em Phys. Rev. E}, 50(4):3092--3099, 1994.

\bibitem{duran96}
J.~Duran, T.~Mazozi, S.~Luding, E.~Cl\'ement, and J.~Rajchenbach.
\newblock Discontinuous decompaction of a falling sandpile.
\newblock {\em Phys. Rev. E}, 53(2):1923, 1996.

\bibitem{luding96}
S.~Luding, J.~Duran, T.~Mazozi, E.~Cl\'ement, and J.~Rajchenbach.
\newblock Simulations of granular flow: Cracks in a falling sandpile.
\newblock In D.~E. Wolf, M.~Schreckenberg, and A.~Bachem, editors, {\em Traffic
  and Granular Flow}, Singapore, 1996. World Scientific.

\bibitem{luding96b}
S.~Luding, J.~Duran, E.~Cl\'ement, and J.~Rajchenbach.
\newblock Simulations of dense granular flow: Dynamic arches and spin
  organization.
\newblock {\em J. Phys. I France}, 6:823--836, 1996.

\bibitem{goldhirsch93}
I.~Goldhirsch and G.~Zanetti.
\newblock Clustering instability in dissipative gases.
\newblock {\em Phys. Rev. Lett.}, 70(11):1619--1622, 1993.

\bibitem{melo94}
F.~Melo, P.~B. Umbanhowar, and H.~L. Swinney.
\newblock Transition to parametric wave patterns in a vertically oscillated
  granular layer.
\newblock {\em Phys. Rev. Lett.}, 72(1):172--175, 1994.

\bibitem{luding96e}
S.~Luding, E.~Cl\'ement, J.~Rajchenbach, and J.~Duran.
\newblock Simulations of pattern formation in vibrated granular media.
\newblock {\em Europhys. Lett.}, 36(4):247--252, 1996.

\bibitem{umbanhowar96}
P.~B. Umbanhowar, F.~Melo, and H.~L. Swinney.
\newblock Localized excitations in a vertically vibrated granular layer.
\newblock {\em Nature}, 382:793--796, 1996.

\bibitem{savage79}
S.~B. Savage.
\newblock Gravity flow of cohesionless granular materials in chutes and
  channels.
\newblock {\em J. Fluid Mech.}, 92:53, 1979.

\bibitem{jenkins79}
J.~T. Jenkins and S.~C. Cowin.
\newblock Theories for flowing granular materials.
\newblock In S.~C. Cowin, editor, {\em Mechanics Applied to the Transport of
  Bulk Materials}, New York, 1979. Am. Soc. Mech. Eng.

\bibitem{haff83}
P.~K. Haff.
\newblock Grain flow as a fluid-mechanical phenomenon.
\newblock {\em J. Fluid Mech.}, 134:401--430, 1983.

\bibitem{homsy92}
G.~M. Homsy, R.~Jackson, and J.~R. Grace.
\newblock Report of a symposium on mechanics of fluidized-beds.
\newblock {\em J. Fluid Mech.}, 236:477, 1992.

\bibitem{hwang95}
H.~Hwang and K.~Hutter.
\newblock A new kinetic model for rapid granular flow.
\newblock {\em Continuum Mech. Thermodyn.}, 7:357--384, 1995.

\bibitem{goldshtein95}
A.~Goldshtein and M.~Shapiro.
\newblock Mechanics of collisional motion of granular materials. {Part 1.
  G}eneral hydrodynamic equations.
\newblock {\em J. Fluid Mech.}, 282:75--114, 1995.

\bibitem{luding98c}
S.~Luding.
\newblock Collisions \& contacts between two particles.
\newblock In H.~J. Herrmann, J.-P. Hovi, and S.~Luding, editors, {\em Physics
  of dry granular media - NATO ASI Series E350}, page 285, Dordrecht, 1998.
  Kluwer Academic Publishers.

\bibitem{brey96}
J.~Javier Brey, M.~J. Ruiz-Montero, and D.~Cubero.
\newblock Homogeneous cooling state of a low-density granular flow.
\newblock {\em Phys. Rev. E}, 54:3664--3671, 1996.

\bibitem{dufty96}
J.~W. Dufty and A.~Santos.
\newblock Practical kinetic model for hard sphere dynamics.
\newblock {\em Phys. Rev. Lett.}, 77(7):1270--1273, 1996.

\bibitem{allen87}
M.~P. Allen and D.~J. Tildesley.
\newblock {\em Computer Simulation of Liquids}.
\newblock Oxford University Press, Oxford, 1987.

\bibitem{lubachevsky91}
B.~D. Lubachevsky.
\newblock How to simulate billards and similar systems.
\newblock {\em J. of Comp. Phys.}, 94(2):255, 1991.

\bibitem{cundall79}
P.~A. Cundall and O.~D.~L. Strack.
\newblock A discrete numerical model for granular assemblies.
\newblock {\em G\'eotechnique}, 29(1):47--65, 1979.

\bibitem{rapaport95}
D.~C. Rapaport.
\newblock {\em The Art of Molecular Dynamics Simulation}.
\newblock Cambridge University Press, Cambridge, 1995.

\bibitem{luding97c}
S.~Luding.
\newblock Surface waves and pattern formation in vibrated granular media.
\newblock In {\em Powders \& Grains 97}, pages 373--376, Amsterdam, 1997.
  Balkema.

\bibitem{luding98f}
S.~Luding and S.~McNamara.
\newblock How to handle the inelastic collapse of a dissipative hard-sphere gas
  with the {TC} model.
\newblock {\em Granular Matter}, 1(3):???, 1998.
\newblock cond-mat/9810009.

\bibitem{johnson89}
K.~L. Johnson.
\newblock {\em Contact Mechanics}.
\newblock Cambridge Univ. Press, Cambridge, 1989.

\bibitem{luding94c}
S.~Luding, H.~J. Herrmann, and A.~Blumen.
\newblock Scaling behavior of 2-dimensional arrays of beads under external
  vibrations.
\newblock {\em Phys. Rev. E}, 50:3100, 1994.

\bibitem{walton86}
O.~R. Walton and R.~L. Braun.
\newblock Viscosity, granular-temperature, and stress calculations for shearing
  assemblies of inelastic, frictional disks.
\newblock {\em Journal of Rheology}, 30(5):949--980, 1986.

\bibitem{foerster94}
S.~F. Foerster, M.~Y. Louge, H.~Chang, and K.~Allia.
\newblock Measurements of the collision properties of small spheres.
\newblock {\em Phys. Fluids}, 6(3):1108--1115, 1994.

\bibitem{labous97}
L.~Labous, A.~D. Rosato, and R.~Dave.
\newblock Measurements of collision properties of spheres using high-speed
  video analysis.
\newblock {\em Phys. Rev. E}, 56:5715, 1997.

\bibitem{luding95b}
S.~Luding.
\newblock Granular materials under vibration: Simulations of rotating spheres.
\newblock {\em Phys. Rev. E}, 52(4):4442, 1995.

\bibitem{luding98b}
S.~Luding, M.~M\"uller, and S.~McNamara.
\newblock The validity of ``molecular chaos'' in granular flows.
\newblock In {\em World Congress on Particle Technology}, Davis Building,
  165-189 Railway Terrace, Rugby CV21 3HQ, UK, 1998. Institution of Chemical
  Engineers.
\newblock ~ISBN 0-85295-401-9.

\bibitem{bernu90}
B.~Bernu and R.~Mazighi.
\newblock One-dimensional bounce of inelastically colliding marbles on a wall.
\newblock {\em J. Phys. A: Math. Gen.}, 23:5745, 1990.

\bibitem{mcnamara92}
S.~McNamara and W.~R. Young.
\newblock Inelastic collapse and clumping in a one-dimensional granular medium.
\newblock {\em Phys. Fluids A}, 4(3):496, 1992.

\bibitem{mcnamara93}
S.~McNamara and W.~R. Young.
\newblock Kinetics of a one-dimensional granular medium in the quasielastic
  limit.
\newblock {\em Phys. Fluids A}, 5(1):34, 1993.

\bibitem{luding94}
S.~Luding, E.~Cl\'ement, A.~Blumen, J.~Rajchenbach, and J.~Duran.
\newblock Studies of columns of beads under external vibrations.
\newblock {\em Phys. Rev. E}, 49(2):1634, 1994.

\bibitem{du95}
Y.~Du, H.~Li, and L.~P. Kadanoff.
\newblock Breakdown of hydrodynamics in a one-dimensional system of inelastic
  particles.
\newblock {\em Phys. Rev. Lett.}, 74(8):1268--1271, 1995.

\bibitem{giese96}
G.~Giese and A.~Zippelius.
\newblock Collision properties of one-dimensional granular particles with
  internal degrees of freedom.
\newblock {\em Phys. Rev. E}, 54:4828, 1996.

\bibitem{deltour97}
P.~Deltour and J.-L. Barrat.
\newblock Quantitative study of a freely cooling granular medium.
\newblock {\em J. Phys. I France}, 7:137--151, 1997.

\bibitem{grossman97}
E.~L. Grossman, T.~Zhou, and E.~Ben-Naim.
\newblock Towards granular hydrodynamics in two-dimensions.
\newblock {\em Phys. Rev. E}, 55:4200, 1997.

\bibitem{shattuck97}
M.~D. Shattuck, C.~Bizon, P.~B. Umbanhowar, J.~B. Swift, and H.~L. Swinney.
\newblock 2d vertically vibrated granular media: Experiment and simulation.
\newblock In {\em Powders \& Grains 97}, Rotterdam, 1997. Balkema.

\bibitem{aspelmeier98}
T.~Aspelmeier, G.~Giese, and A.~Zippelius.
\newblock Cooling dynamics of a dilute gas of inelastic rods: A many particle
  simulation.
\newblock {\em Phys. Rev. E}, 57:857, 1998.

\bibitem{luding94d}
S.~Luding, E.~Cl\'ement, A.~Blumen, J.~Rajchenbach, and J.~Duran.
\newblock Anomalous energy dissipation in molecular dynamics simulations of
  grains: The ``detachment effect''.
\newblock {\em Phys. Rev. E}, 50:4113, 1994.

\bibitem{luding95}
S.~Luding, E.~Cl\'ement, A.~Blumen, J.~Rajchenbach, and J.~Duran.
\newblock Interaction laws and the detachment effect in granular media.
\newblock In {\em Fractal Aspects of Materials}, volume 367, pages 495--500,
  Pittsburgh, Pennsylvania, 1995. Materials Research Society, Symposium
  Proceedings.

\bibitem{landau75}
L.~D. Landau and E.~M. Lifshitz.
\newblock {\em Elasticity Theory}.
\newblock Pergamon Press, Oxford, 1975.

\bibitem{kuwabara87}
G.~Kuwabara and K.~Kono.
\newblock Restitution coefficient in a collision between two spheres.
\newblock {\em Japanese Journal of Applied Physics}, 26(8):1230--1233, 1987.

\bibitem{brilliantov96}
N.~V. Brilliantov, F.~Spahn, J.~M. Hertzsch, and T.~P\"oschel.
\newblock Model for collisions in granular gases.
\newblock {\em Phys. Rev. E}, 53(5):5382, 1996.

\bibitem{lee93}
J.~Lee and H.~J. Herrmann.
\newblock Angle of repose and angle of marginal stability: Molecular dynamics
  of granular particles.
\newblock {\em J. Phys. A}, 26:373, 1993.

\bibitem{schafer96}
J.~Sch\"afer, S.~Dippel, and D.~E. Wolf.
\newblock Force schemes in simulations of granular materials.
\newblock {\em J. Phys. I France}, 6:5--20, 1996.

\bibitem{radjai97c}
F.~Radjai, J.~Sch\"afer, S.~Dippel, and D.~Wolf.
\newblock Collective friction of an array of particles: A crucial test for
  numerical algorithms.
\newblock {\em J. Phys. I France}, 7:1053, 1997.

\bibitem{poschel93}
T.~P\"oschel and V.~Buchholtz.
\newblock Static friction phenomena in granular materials: Coulomb law vs.
  particle geometry.
\newblock {\em Phys. Rev. Lett.}, 71(24):3963, 1993.

\bibitem{walton93c}
O.~R. Walton and R.~L. Braun.
\newblock Simulation of rotary-drum and repose tests for frictional spheres and
  rigid sphere clusters.
\newblock In {\em DOE/NSF Workshop on Flow of Particulates and Fluids}, pages
  1--17, 1993.

\bibitem{walton94}
O.~R. Walton.
\newblock Effects of interparticle friction and particle shape on dynamic
  angles of repose via particle-dynamics simulation.
\newblock In {\em Workshop: Mechanics and Statistical Physics of Particulate
  Materials}, 1994.

\bibitem{poschel95}
T.~P\"oschel and V.~Buchholtz.
\newblock Molecular dynamics of arbitrarily shaped granular particles.
\newblock {\em J. Phys. I France}, 5(11):1431--1455, 1995.

\bibitem{buchholtz94}
V.~Buchholtz and T.~P\"oschel.
\newblock Numerical investigations of the evolution of sandpiles.
\newblock {\em Physica A}, 202:390, 1994.

\bibitem{buchholtz95}
V.~Buchholtz, T.~P\"oschel, and H.-J. Tillemans.
\newblock Simulation of rotating drum experiments using non-circular particles.
\newblock {\em Physica A}, 216:199, 1995.

\bibitem{kohring95}
G.~A. Kohring, S.~Melin, H.~Puhl, H.~J. Tillemans, and W.~Verm\"ohlen.
\newblock Computer simulations of critical, non-stationary granular flow
  through a hopper.
\newblock {\em Comput. Methods in Appl. Mechanics and Eng.}, 124:2273, 1995.

\bibitem{matuttis97}
H.-G. Matuttis and S.~Luding.
\newblock The effect of particle shape and friction on the stresses in heaps of
  granular media.
\newblock In D.~E. Wolf and P.~Grassberger, editors, {\em Friction, Arching and
  Contact Dynamics}, Singapore, 1997. World Scientific.

\bibitem{brendel98}
L.~Brendel and S.~Dippel.
\newblock Lasting contacts in molecular dynamics simulations.
\newblock In H.~J. Herrmann, J.-P. Hovi, and S.~Luding, editors, {\em Physics
  of Dry Granular Media}, page 313, Dordrecht, 1998. Kluwer Academic
  Publishers.

\bibitem{bird94}
G.~A. Bird.
\newblock {\em Molecular Gas Dynamics and the Direct Simulation of Gas Flows}.
\newblock Clarendon, Oxford, 1994.

\bibitem{tanaka96}
T.~Tanaka, S.~Yonemura, K.~Kiriba Yashi, and Y.~Tsuji.
\newblock Cluster formation and particle-induced instability in gas-solid flows
  predicted by the dsmc method.
\newblock {\em JSME Int. Journal B}, 39(2):239--245, 1996.

\bibitem{muller97}
M.~M\"uller, S.~Luding, and H.~J. Herrmann.
\newblock Simulations of vibrated granular media in 2d and 3d.
\newblock In D.~E. Wolf and P.~Grassberger, editors, {\em Friction, Arching and
  Contact Dynamics}, Singapore, 1997. World Scientific.

\bibitem{gervois92}
A.~Gervois and D.~Bideau.
\newblock Some geometrical properties of two-dimensional hard disk packings.
\newblock In D.~Bideau, editor, {\em Disorder and Granular Media}, Amsterdam,
  1992. North Holland.

\bibitem{alexander95b}
F.~J. Alexander, A.~L. Garcia, and B.~J. Alder.
\newblock Simulation of the consistent {B}oltzmann equation for hard spheres
  and its extension to dense gases.
\newblock In {\em Lecture Notes in Physics}, Berlin, 1995. Springer Verlag.

\bibitem{goldhirsch93b}
I.~Goldhirsch, M.-L. Tan, and G.~Zanetti.
\newblock A molecular dynamical study of granular fluids {I: T}he unforced
  granular gas in two dimensions.
\newblock {\em Journal of Scientific Computing}, 8:1--40, 1993.

\bibitem{mcnamara96}
S.~McNamara and W.~R. Young.
\newblock Dynamics of a freely evolving, two-dimensional granular medium.
\newblock {\em Phys. Rev. E}, 53(5):5089--5100, 1996.

\bibitem{jenkins85b}
J.~T. Jenkins and M.~W. Richman.
\newblock Kinetic theory for plane flows of a dense gas of identical, rough,
  inelastic, circular disks.
\newblock {\em Phys. of Fluids}, 28:3485--3494, 1985.

\bibitem{luding98d}
S.~Luding, M.~Huthmann, S.~McNamara, and A.~Zippelius.
\newblock Homogeneous cooling of rough dissipative particles: Theory and
  simulations.
\newblock {\em Phys. Rev. E}, 58:3416--3425, 1998.

\bibitem{grossman96b}
E.~L. Grossman and B.~Roman.
\newblock Density variations in a one-dimensional granular system.
\newblock {\em Phys. Fluids}, 8:3218, 1996.

\bibitem{kudrolli97}
A.~Kudrolli and J.~P. Gollub.
\newblock Studies of cluster formation due to collisions in granular material.
\newblock In {\em Powders \& Grains 97}, page 535, Rotterdam, 1997. Balkema.

\bibitem{kudrolli97b}
A.~Kudrolli, M.~Wolpert, and J.P. Gollub.
\newblock Cluster formation due to collisions in granular material.
\newblock {\em Phys. Rev. Lett.}, 78(7):1383--1386, 1997.

\bibitem{sibuya90}
M.~Sibuya, T.~Kawai, and K.~Shida.
\newblock Equipartition of particles forming clusters by inelastic collisions.
\newblock {\em Physica A}, 167:676, 1990.

\bibitem{mcnamara94}
S.~McNamara and W.~R. Young.
\newblock Inelastic collapse in two dimensions.
\newblock {\em Phys. Rev. E}, 50(1):R28--R31, 1994.

\bibitem{trizac95}
E.~Trizac and J.~P. Hansen.
\newblock Dynamic scaling behavior of ballistic coalescence.
\newblock {\em Phys. Rev. Lett.}, 74(21):4114--4117, 1995.

\bibitem{spahn97}
F.~Spahn, U.~Schwarz, and J.~Kurths.
\newblock Clustering of granular assemblies with temperature dependent
  restitution and under keplerian differential rotation.
\newblock {\em Phys. Rev. Lett.}, 78:1596--1599, 1997.

\bibitem{orza97}
J.~A.~G. Orza, R.~Brito, T.~P.~C. van Noije, and M.~H. Ernst.
\newblock Patterns and long range correlations in idealized granular flows.
\newblock {\em Int. J. of Mod. Phys. C}, 8:953, 1997.

\bibitem{trizac96}
E.~Trizac and J.~P. Hansen.
\newblock Dynamics and growth of particles undergoing ballistic coalescence.
\newblock {\em J. Stat. Phys.}, 82:1345--1370, 1996.

\bibitem{mcnamara96b}
S.~McNamara.
\newblock Two dimensional granular medium driven by a vibrating wall.
\newblock preprint, 1996.

\bibitem{goldhirsch96}
I.~Goldhirsch and M.-L. Tan.
\newblock The single-particle distribution function for rapid granular shear
  flows of smooth inelastic disks.
\newblock {\em Phys. Fluids}, 8(7):1752--1763, 1996.

\bibitem{duran94}
J.~Duran, T.~Mazozi, E.~Cl\'ement, and J.~Rajchenbach.
\newblock Size segregation in a two-dimensional sandpile: convection and
  arching effects.
\newblock {\em Phys. Rev. E}, 50(6):5138--5141, 1994.

\bibitem{luding96c}
S.~Luding, J.~Duran, E.~Cl\'ement, and J.~Rajchenbach.
\newblock Computer simulations and experiments of dry granular media:
  Polydisperse disks in a vertical pipe.
\newblock In {\em Proc. of the 5th Chemical Engineering World Congress}, San
  Diego, 1996. AIChE.

\end{thebibliography}

\end{document}